\documentclass[12pt]{article}
\pdfoutput=1
\usepackage{graphics, color}
\usepackage{graphicx}

\usepackage[pdftex,colorlinks=true,linkcolor=blue,citecolor=blue]{hyperref}
\usepackage{enumerate}

\usepackage{mathrsfs}

\usepackage{caption}
\usepackage{subcaption}
\usepackage{amssymb}
\usepackage{amsmath}
\usepackage{cite}
\pdfoutput=1

\newcommand{\figref}[1]{Fig. \ref{#1}}

\def\gsim{\, \rlap{$>$}{\lower 1.1ex\hbox{$\sim$}}\,}
\def\lsim{\, \rlap{$<$}{\lower 1.1ex\hbox{$\sim$}}\,}

 \def\simleq{\; \raise0.3ex\hbox{$<$\kern-0.75em
      \raise-1.1ex\hbox{$\sim$}}\; }
   \def\simgeq{\; \raise0.3ex\hbox{$>$\kern-0.75em
      \raise-1.1ex\hbox{$\sim$}}\; }

\def\Op{{\mathcal{O}}}

 \newcommand{\be}{\begin{equation}}
\newcommand{\ee}{\end{equation}}
 \newcommand{\bal}{\begin{align}}
 \newcommand{\eal}{\end{align}}
\newcommand{\ben}{\begin{equation*}}
\newcommand{\een}{\end{equation*}}
\newcommand{\bea}{\begin{eqnarray}}
\newcommand{\eea}{\end{eqnarray}}
\newcommand{\bean}{\begin{eqnarray*}}
\newcommand{\eean}{\end{eqnarray*}}
\newcommand{\bes}{\begin{subequations}}
\newcommand{\ees}{\end{subequations}}

\def\p{\partial}

\def\eps{{\epsilon}}

\usepackage{epsfig}

\numberwithin{equation}{section}

\begin{document}


\begin{titlepage}

\begin{flushright}
BRX-TH-671\\
EFI-13-31
\end{flushright}

\bigskip
\bigskip\bigskip\bigskip
\centerline{\Large \bf Metastability and instability }
\smallskip
\centerline{\Large \bf in holographic gauge theories}
\bigskip\bigskip\bigskip

\centerline{{\bf Matthew Kleban${}^{1}$, Albion Lawrence${}^{2}$,}}
\centerline{{\bf Matthew M. Roberts${}^{3}$, and Stefano Storace${}^{1}$}}
\medskip
\centerline{\em ${}^1$ Center for Cosmology and Particle Physics}
\centerline{\em Department of Physics, New York University}
\centerline{\em 4 Washington Place, New York, NY 10003, and }
\centerline{\em      New York University Abu Dhabi, United Arab Emirates}
\medskip
\smallskip
\centerline{\em ${}^2$ Martin Fisher School of Physics, Brandeis University}
\centerline{\em MS 057, 415 South Street, Waltham, MA 02454}
\medskip
\centerline{\em ${}^3$ Kadanoff Center for Theoretical Physics and Enrico Fermi Institute,}
\centerline{\em The University of Chicago, 5640 South Ellis Ave., Chicago, IL 60637}
\bigskip
\bigskip\bigskip


\begin{abstract}

We review and extend previous results regarding the stability and thermodynamics of Anti-de Sitter (AdS) spacetime at finite temperature.  Using a combination of analytic and numerical techniques, we compute the energy, temperature, and entropy of perfect fluid stars in asymptotically AdS spacetimes.  We find that at sufficiently high  temperature (in the canonical ensemble) or energy (in the microcanonical ensemble) these configurations develop dynamical instabilities, which presumably lead to the formation of a black hole.   We extend our analysis to the case of $AdS \times S$  compactifications stabilized by flux (such as those that arise in supergravity and string theory), and find that the inclusion of the sphere does not substantially alter these results.  We then map out the phase structure of these theories in the canonical and microcanonical ensembles, paying attention to inequivalence of these ensembles for global anti-de Sitter space.   With a certain scaling limit, the critical temperature can be parametrically lower than the string temperature, so that supergravity is a good description at the instability point.  We comment on the implications of this for the unitarity of black holes.

\end{abstract}
\end{titlepage}
\baselineskip = 17pt
\tableofcontents
\setcounter{footnote}{0}

\section{Introduction}

Theories described by Einstein gravity plus a few massless fields (such as those appearing in supergravity), in asymptotically anti-de Sitter spacetimes, have a well-known first-order phase transition at temperatures of order the inverse curvature radius of the background AdS spacetime \cite{Hawking:1982dh}.  At this ``Hawking-Page'' temperature $T_{HP}$, the solutions corresponding to a black hole with AdS-scale curvature and to a self-gravitating hot gas of massless particles exchange dominance in the free energy landscape.  When such theories appear as holographic duals to large-N, strongly coupled gauge theories, they typically signal a deconfinement transition with a large jump in entropy, typically a deconfinement transition \cite{Witten:1998qj,Witten:1998zw}.

Whether one is studying gravity, gauge theory, or the relationship between them, the thermodynamic landscape in various ensembles -- the maxima, minima and saddle points of the entropy and of various free energies -- contains important additional information.  In a first order transition, metastable phases provide important long-time intermediate phases when cooling the system, and can appear as transients in the long-time response to a perturbation.  For example, in holographic gauge theories, the contribution of the metastable supergravity gas to the long-time behavior of the bulk finite-temperature dynamics above the Hawking-Page transition has been argued to be a key to understanding unitarity in the quantum dynamics of black holes \cite{Maldacena:2001kr}. 

Another interesting configuration is the black hole with sub-AdS-scale horizon, which has negative specific heat; it is unstable in the canonical ensemble, but potentially stable in the microcanonical ensemble, for a range of masses \cite{Hawking:1982dh,Horowitz:1999uv}.  Such black holes should provide a key to understanding holography at sub-AdS scales.

In this work we will attempt to provide a unified overview of the thermal landscape of simple theories of gravity coupled to light particles, the latter modeled as a self-gravitating fluid.  In \S2\ we will study static configurations as a function of core density, and compute their energy $E$, temperature $T$, entropy $S$, and Helmholtz free energy $F = E - TS$. For fluids in asymptotically $AdS_D$ spacetimes, these have been found and the first three quantities computed \cite{Page:1985em,Hubeny:2006yu,Vaganov:2007at,Hammersley:2007rp}; we will review those solutions for $D=3,5$ and compute in addition the free energy $F(T)$ and the specific heat at constant volume $C_V$.    We will study the stability to perturbations at fixed energy and temperature.  For self-gravitating gasses, the instabilities for fixed-energy perturbations were identified based on the assumption that solutions which were local maxima of the entropy were stable to local perturbations of the fluid density \cite{Page:1985em,Vaganov:2007at,Hammersley:2007rp}.  This is a proven theorem for asymptotically flat spactimes \cite{Sorkin:1981wd}, but not for asymptotically AdS spacetimes.  Our results are consistent with the conjecture that this theorem extends to asymptotically AdS spacetimes (other work on the stability of stars in AdS includes \cite{deBoer:2009wk,Hartnoll:2010gu,Hartnoll:2010ik}).  This story is closely related to recent work on the stability properties of black holes \cite{Gubser:2000ec,Gubser:2000mm,Hubeny:2002xn,Friess:2005zp,Hollands:2012sf}.  The results for fixed temperature are new.  They give the unsurprising result that the instability for temperature-preserving perturbations sets in when the specific heat turns negative, but also show that the instability persists at higher core density as the specific heat turns positive. Finally, in the $AdS_3$ and $AdS_{5}$ cases we consider the inclusion of  $S_5$ and $S_{3}$ factors in a Freund-Rubin compactification of type IIB supergravity, and show that this does not significantly affect the physics (although it does make the numerical solutions more difficult to extract).  Our analysis is restricted to configurations that are homogeneous on the $S_{D}$, but for the general reasons explained in Sec.~\ref{reasons} we do not expect inhomogeneous solutions (if they exist) to alter our main conclusions.

In \S3\ we will combine the rich landscape of static configurations of \S2\ with the landscape of black hole solutions, to discuss the thermodynamic landscape of gauge theories with holographic AdS duals. One lesson, perhaps understood by many but not spelled out clearly in the literature, is that the landscape and the pattern of instabilities differ in the canonical and microcanonical ensembles.  It is known that for quantum theories in finite volume, and for theories with sufficiently long-range interactions, the canonical and microcanonical ensembles are not equivalent. This allows for the existence of stable configurations in the microcanonical ensemble -- small black holes -- with negative specific heat.  We will review this story of ensemble inequivalence, including comparatively recent results regarding the relationship between them \cite{touchette2005nonequivalent}, and discuss the microcanonical and canonical phase diagrams of holographic gauge theories in this framework. 

Finally, the implication of our work is that there is a maximum mass above which the self-gravitating gas of radiation simply ceases to be a solution. While this was known to happen at temperatures of order the string scale \cite{Kruczenski:2005pj}, we find that in a certain large-N scaling limit, the instability sets in at lower temperatures.  As we will argue, this implies that the gas of light supergravity modes cannot explain the long-time behavior of finite-temperature correlation functions, contrary to the conjecture in \cite{Maldacena:2001kr}.

\section{Self-gravitating hot gasses in $AdS_{D}$ and $AdS_{D}\times S_{D}$}

In this section we will describe the classical static configurations of gravitating hot fluids coupled to gravity in asymptotically $AdS_D$ backgrounds, and study their stability properties.  This will allow us to give a fuller picture of the thermodynamic landscape in \S3.  We will find via explicit calculation in $AdS_3$ and numerical checks in $AdS_5$ that perturbations which fix the energy go unstable precisely when the entropy as a function of the total mass (or core density) ceases to be a local maximum.  This is known to be true in asymptotically flat four-dimensional geometries \cite{Sorkin:1981wd}. While it was assumed that this relationship between local maximization of the entropy and stability would also hold for asymptotically AdS spacetimes by \cite{Page:1985em,Hammersley:2007rp,Vaganov:2007at}, to our knowledge our results provide the first check of this assumption (which remains to be proven).

We will also examine the solutions for fluids in $AdS_5\times S_5$ and $AdS_3\times S_3$; the essential point will be that while the equations of motion are more difficult, the presence of the  sphere does not sigificantly alter the solutions we found in pure AdS.  We will also provide a hand-waving argument that the initial instability for the self-gravitating hot fluid is homogeneous on the $S_D$ factor of the spacetime.  However, this does not exclude the possibility that the endpoint of the homogeneous instability is a meta-stable configuration that is localized on $S_D$.  Even if such configurations exist, however, we expect them to become unstable at some higher temperature for the general reasons given below.

Before studying the full solution, we will review the classic argument for the instability of gravitating systems at finite temperature, and provide a parametric bound for the onset of such an instability in asymptotically AdS backgrounds.

\subsection{Stability and graviton self-energy for hot spacetimes}\label{reasons}

Before entering into a full analysis of the solution to Einstein's equations, we recall prior work regarding the stability of flat and asymptotically AdS spacetimes in the presence of a fluid with sufficiently high density or pressure.  Although it is not consistent to ignore backreaction, this analysis does give the correct parametric dependence of the onset of instability as a function of the density or temperature.

The initial observation of the instability of sufficiently uniform matter in Newtonian gravity is due to Jeans \cite{Jeans:1902}. One finds that the linearized density perturbations satisfy a second-order wave equation with an imaginary ``Jeans'' mass depending on the background density.  Perturbations at wavelengths longer than the magnitude of the inverse Jeans mass are unstable and grow exponentially.  Dimensional analysis shows that the Jeans mass $M_J$ must be 
\be
	M_J^2 = - \kappa^2 \rho
\ee
where $\kappa^2$ is Newton's constant (also known as $8\pi G_N$).

A relativistic version of this argument was provided by Gross, Perry, and Yaffe \cite{Gross:1982cv}\ in their study of hot flat space.  The authors showed that the component $g_{00}$ of the graviton (in the frame at which the thermal bath is at rest) obtains a tachyonic mass term at one loop, 
\be
	\delta m^2 (g_{00}) \sim - \kappa^2 \cdot T^{D} \ ,
	\label{gravtachyon}
\ee
where $T$ is the temperature, $\kappa^2$ the gravitational coupling (with dimension (length)${}^{D-2}$), and we have generalized the result to $D$ spacetime dimensions.  This again essentially follows from dimensional analysis, and is in keeping with the above result if $\rho = \sigma T^D$ for some Stefan-Boltzmann constant $\sigma$.

The calculation of \cite{Gross:1982cv} is not easily generalized to the curved space backgrounds we are interested in.  Furthermore, it has the serious flaw that the hot fluid in flat space is  not a solution to Einstein's equations, so there will be a ``tadpole'' term in the graviton equations of motion at the same order as the tachyon term.  More generally, the canonical ensemble is not well defined for self-gravitating systems in asymptotically flat spacetimes.\footnote{Of course, in the early universe the finite temperature radiation and matter source expansion, and a notion of local thermal equilibrium is still available. At distances below the cosmological horizon scale, the calculations above are quite relevant for understanding the formation of structure in the linear regime.}  Nonetheless, it gives the correct parametric dependence of the onset of instability when that instabilty sets in at a wavelength below the radius of curvature of the correct spacetime.

We will be interested in asymptotically anti-de Sitter backgrounds, which provide a kind of gravitational ``box'' inside of which thermodynamics for gravitating systems can be made sensible.  The negative background curvature changes the stability, as
a scalar with negative (mass)${}^2$ can remain stable provided the squared mass is above the Breitenlohner-Freedman bound \cite{1982AnPhy.144..249B}:
\begin{equation}
 m^2 \geq -\frac{D^2}{4L^2} \ ,
\label{BFbound}
\end{equation}
where $L$ is the AdS length.  The parametric dependence agrees with the tentative conclusion of work by Gribosky, Donoghue and Holstein \cite{Gribosky1989149} who, using real time formalism for quantum field theory at finite temperature and accounting for the tadpole generated by the classical gravitational backreaction, generalized \cite{Gross:1982cv} by expanding in powers of the spacetime curvature.   Their result was that there is indeed a negative self-energy of the form \eqref{gravtachyon}, although with a different ${\cal O}(1)$ coefficient than that of \cite{Gross:1982cv}.  

Eqn (\ref{BFbound}) together with the results of \cite{Gross:1982cv} suggest that
an instability, if any, should occur at temperatures of order 
\be
	T_J^D \sim \frac{1}{\kappa_D^2 L^2}
\ee
In the limit that the radius of curvature is below the Planck scale, this temperature is well above $1/L$.  Note that for typical Freund-Rubin compactifications on $AdS_p\times S_q$, the radius of curvature of both factors is of the same order $L$, and we should set $D = p + q$.  Furthermore, if there are additional factors in the compactification geometry whose radius of curvature is $\gtrsim T_J^{-1}$, these will also alter the story.

Our arguments above give a crude scaling; they do not take into account the details of the effects of the AdS background on the density profile of thermally excited matter, or the backreaction of said matter on the metric.  Therefore, we will proceed to directly studying the spacetime background corresponding to a self-gravitating fluid.  While the above estimate for the temperature at which an instability sets in is correct, the full story is much richer.

\subsection{General equations for $AdS_D$ backgrounds} \label{subsec:gensetup}

\subsubsection{Static solutions}

At finite temperature,  spacetime is filled with a gas of particles in thermal equilibrium. The evolution of a $D = d+1$-dimensional system with AdS length $L$ is therefore described by Einstein's equations coupled to the stress-energy tensor of thermal matter:
\be
R_{ab} - g_{ab} R/2 - d(d-1)/2L^2 g_{ab}= \kappa^2 T^M_{ab},\label{eq:EinEq}
\ee
with the usual conventions. We approximate the gas as a perfect fluid 
described by a stress-energy tensor of the form
\be
T^{ab} = (\rho+p)u^au^b+p g^{ab},\label{fluidtensor}
\ee
where $\rho$ and $p$ are the density and the pressure, respectively, and 
$u^a = \partial x^a / \partial_\tau$ is the flow vector of the fluid.
We will specialize to spherically symmetric solutions, with metric  ans\"atz
\be
ds^2 = -f(r)e^{-\chi(r)}dt^2 + dr^2/f(r) + r^2 d\Omega_{d-1}, \label{ansatz}
\ee
where $d\Omega_{d-1}$ is the metric of a $d-1$-dimensional sphere.

A complete specification of the dynamics requires an equation of state $p = p(\rho)$.  As in the case for static self-gravitating fluids in asymptotically flat spacetimes \cite{Tolman:1939jz,Oppenheimer:1939ne}, Einstein's equations, together with the conservation of stress-energy $\nabla_a T^{ab}=0$, give rise to a set of {\it first} order equations of motion:
\be
f p_{,\rho}\rho'+(\rho+p)\left(\frac{dr}{2L^2}+\frac{\kappa^2 r }{d-1}p+\frac{(d-2)(1-f)}{2r} \right)=0,\label{eq:bg_dens}
\ee
\be
f'+\frac{2-d}{r}-\frac{d r}{L^2}+\frac{(d-2)f}{r}+\frac{2\kappa^2 r }{d-1}\rho=0,\label{eq:bg_f}
\ee
\be
\chi'+\frac{2\kappa^2 r}{(d-1)f}\left(\rho+p \right)=0,\label{eq:bg_chi}
\ee
where the prime sign stands for derivative with respect to $r$.  We should also include the equation of state $p(\rho)$; from here on out, we will  assume a linear equation of state $p=w \rho$.\footnote{In Appendix \ref{app:full_pert} we present equations of motion for perturbations with a general equation of state.}  We expect this to be a good approximation so long as the massless supergravity modes are the dominant component of the fluid, as these should behave like radiation with $w=(D-1)^{-1}$. Moreover, the primary interactions are dimensionful, suppressed by the string or Planck scale.  Since we are studying sub-stringy temperatures and densities, we expect microscopic interactions, and therefore the consequent changes in the equation of state, to be small. At sufficiently high temperatures  Kaluza-Klein modes on the sphere will play a role and this approximation might break down, although presumably this simply corresponds to a higher dimensional radiation equation of state.  Nevertheless it would be interesting to study the effect of non-linear equations of state, which we will leave  for future work.

 It appears that three constants of motion are needed, for three first-order equations with unknowns $f,\chi, \rho$.  We demand that the solution be asymptotically anti-de Sitter, so that as $r \to \infty$, $f \to r^2/L^2 + {\cal O}(1)$ and $\chi \rightarrow 0$.  (Note that only the first derivative of $\chi$ appears, so that a constant term is unfixed by these equations; this can be shifted by rescaling $t$, so it is an unphysical constant).  We demand regularity at the origin, leaving either the mass term at infinity (the coefficient of $- 1/r^{d-2}$) or the ``core density'' $\rho_0\equiv \rho(r=0)$, at the origin as the constant of motion.  

\begin{figure}[thbp]
\begin{center}
\includegraphics[scale=.5]{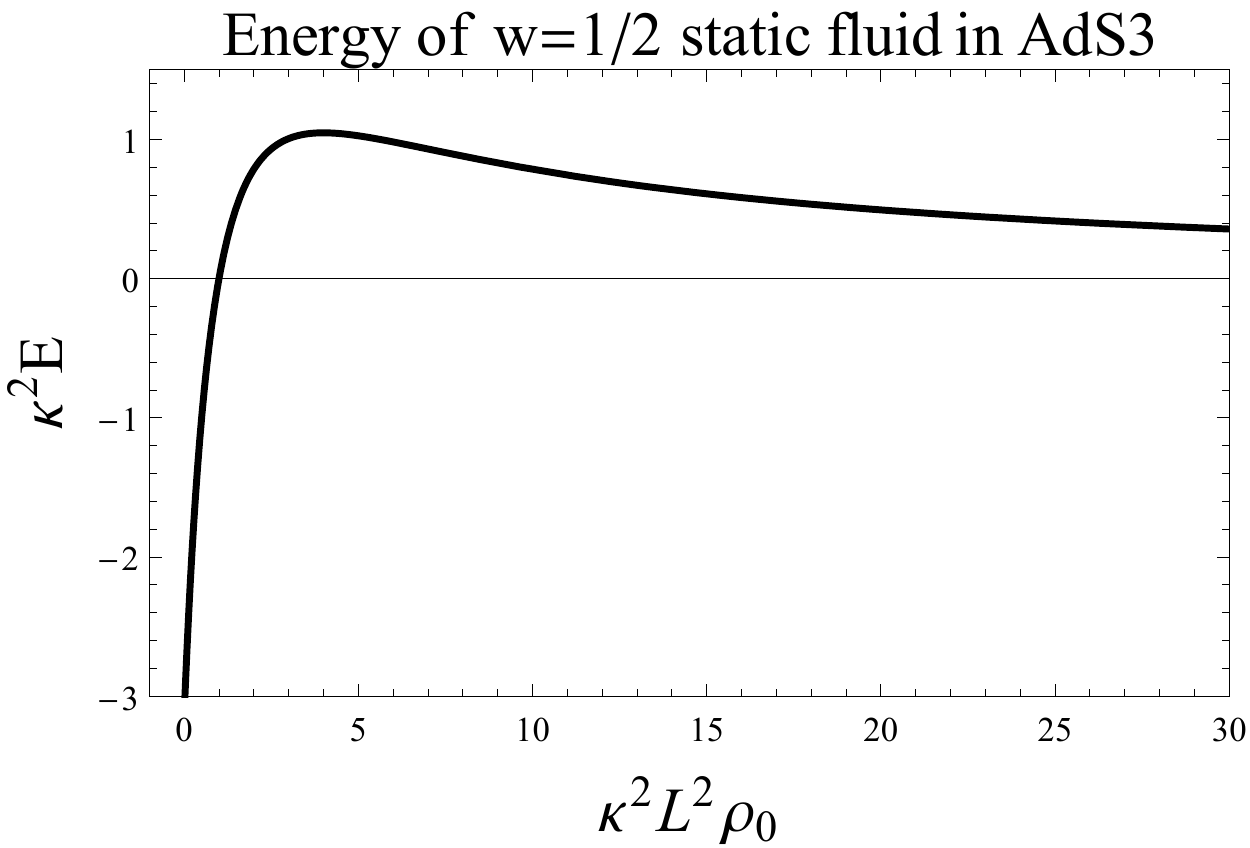}
\includegraphics[scale=.5]{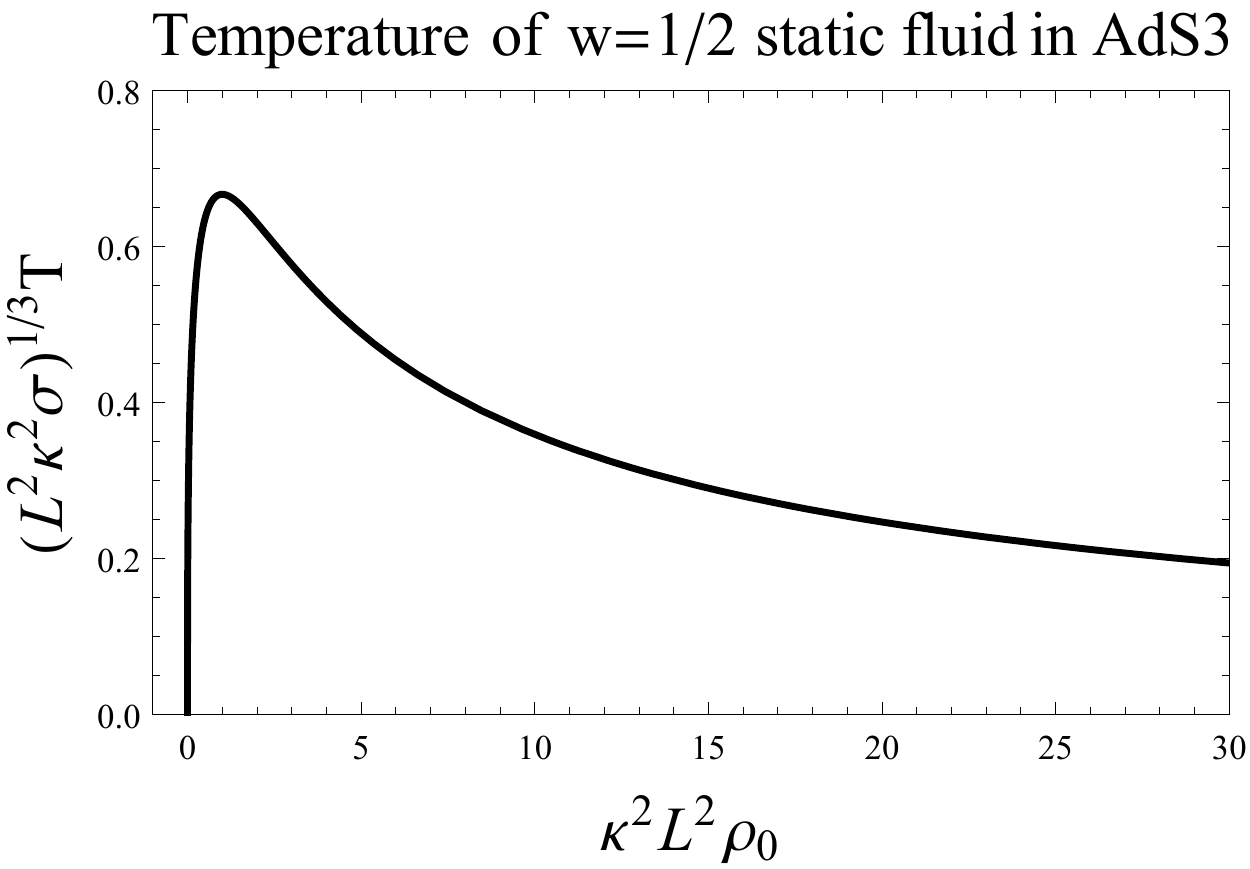}
\caption{The total energy $E$ and temperature $T$
of spherical stars in $AdS_3$ for radiation ($w=1/2$), as functions
of the core density $\rho_{0}$.
\label{fig:ads3masstemp}}
\end{center}
\end{figure}

In practice, we will choose the core density $\rho_0$, integrating out from the origin.  Global thermodynamic quantities 
such as the energy and temperature can be computed as functions of the core density.  They are not in general monotonic, but 
instead reach a maximum and then decrease (due to the large gravitational backreaction at high densities).  To give a preview, in   \figref{fig:ads3masstemp}\ we plot as a representative case the energy and boundary temperature as functions of the core density for the case of a radiation ($w=1/2$) gas in $AdS_{3}$ (we will describe how we extract these quantities below), based on the analytic solutions in \cite{Garcia:2002rn}.  The temperature and energy both have a maximum value, with the maximum temperature occurring at a value of $\rho_{0,T_{max}}$ smaller than that for the maximum mass $\rho_{0,E_{max}}$. Thus, for $\rho_{0,T_{max}} < \rho_0 < \rho_{0,E_{max}}$ the specific heat $c_{v} \propto \p M/\p T < 0$, indicating an instability of the canonical ensemble.  For $\rho_{0} > \rho_{0,E_{max}}$  a tachyonic mode appears and the star becomes dynamically unstable\footnote{The latter, or at least the existence of a zero mode at this critical value of $\rho_{0}$, follows from a theorem given in \cite{weinbook}.}. We expect the same qualitative features to persist at least for $0<w<1$ and for a range of dimensionalities, and we will check it explicitly for radiation in $AdS_5$, $AdS_3\times S_3$, 
and $AdS_5\times S_5$.

\subsubsection{Thermodynamic quantities}

Each value of the core density $\rho_0$ corresponds to a distinct geometry with energy $E$ and
entropy $S$ in the microcanonical ensemble, or free energy $F$ and temperature $T$ in the
canonical ensemble. We can eliminate $\rho_0$ and use the curves $S(E)$ and $F(T)$ to describe microcanonical and canonical phases, respectively.  In doing so we need to choose the maximum of $S$ or the minimum of $F$, in the cases where there are several classical solutions for a given $E$ or $T$, as seen in \figref{fig:ads3masstemp}.  

The energy is given by the ADM mass \cite{PhysRev.116.1322} \footnote{For $\frac{1}{d-1}<w<1$ the backreaction of the fluid on the metric leads to divergences in the ADM mass, which can be cancelled by appropriate holographic renormalization. This never occurs for $w\le w_{rad.}=1/d$.}
\be
E\equiv M_{ADM}=\frac{1}{2\kappa^2}\int_{S_\infty}{(g_{ij},_j-g_{jj},_i)dn^i}\ , \label{eq:admformula}
\ee
where latin indices refer to spatial components, repeated indices are summed over, $S_\infty$ is a sphere at infinity
and $n^i$ is a radial unit vector pointing outside $S_\infty$. Given this formula, we can compute $E$ starting from the  ans\"atz.  
Once Eqns. (\ref{eq:bg_dens}-\ref{eq:bg_f}) have been solved for a given value of $\rho_0$, we can then insert the specific values of $f$, $\rho$, and $\chi$ into the ans\"atz (\ref{eq:EinEq}) and then insert that into (\ref{eq:admformula}) to get $E(\rho_0)$.
 
To compute $T$ we assume thermodynamic equilibrium holds, which implies:
\be 
T_{loc} = \sqrt{-g^{tt}} T\ .
\ee
for constant $T$, where $T_{loc}$ is the ``proper temperature'' measured by a physical thermometer.  We take $T$ to be the temperature in the dual field theory.  We will consider fluids with a linear equation of state $p = w \rho$; given $\rho(r)$, we can then calculate $T_{loc}$ via the appropriate generalization of the Stefan-Boltzmann law:
\be \rho =\sigma T_{loc}^\frac{1+w}{w}=\sigma \left(-g_{tt} \right)^{-\frac{1+w}{2w}}T^\frac{1+w}{w}\ .
\ee
At large radius, the local density falls off as $\rho \sim \mu/r^\frac{1+w}{w}$ for some constant $\mu$.  Thus, we can extract the global temperature:
\be
T = (\mu/\sigma)^\frac{w}{1+w}/L\ .\label{eq:w_temp_density}
\ee

To compute the microcanonical entropy $S(E)$, we again use local thermodynamic equilibrium to compute a coarse-grained entropy density given $\rho$, $p = w \rho$, and $T_{loc}$. The total entropy $S$ is the integral over space of the entropy density
\be
s=\frac{p(\rho)+\rho}{T_{loc}}=(1+w)\sigma^\frac{w}{1+w}\rho^\frac{1}{1+w}.
\ee
One can show that for solutions to (\ref{eq:bg_dens}-\ref{eq:bg_f}), with $\rho, T$ determined by local thermal equilibrium, the above definitions of $S,T,E$ are consistent with the first law $dE = T dS$. Finally, the free energy of a given solution is $F=E-TS$.\footnote{Note that we are not assuming that the canonical and microcanonical ensembles are equivalent -- as we will discuss in \S3\ below, they are not. This inequivalence means that we cannot use $F(T)$ to extract $S(E)$, as $T$ is a multivalued function of $E$.} 

\subsubsection{Dynamical stability} \label{subsubsec:mechstab}

We wish to understand the mechanical stability of our solutions under perturbations that preserve either the energy or the temperature, and how such stability is correlated with the thermodynamic properties of these solutions.

We consider small deviations from the the stress-energy tensor (\ref{fluidtensor}) and metric (\ref{ansatz}). For solutions with harmonic time dependence $\sim e^{-i\omega t}$ for all fields (in the linear approximation), a dynamical instability is present when there are imaginary frequencies in the spectrum. The symmetry of the initial configuration suggests that the first  instability will be  an s-wave perturbation, of the form $\delta g_{tt}(t,r),~\delta g_{rr}(t,r),~\delta u^a(t,r),~\delta \rho(t,r)$ and $\delta p(t,r)$. 

The full set of equations for these perturbations can be found in Appendix \ref{app:full_pert}.  Here, we give the equations for the perturbations of the source fields $\delta \rho$ and $\delta u^r$ for the case of a linear equation of state, where the sound speed of perturbations $c_s^2\equiv \delta p / \delta \rho = w$ is constant:
 \begin{align}
\delta\rho'(r)  = &~~~ \left(
\frac{i\kappa^2r^2e^{-\chi/2}(p+\rho)^2}{(d-1)f^{3/2}w}\left\{ 
\frac{d}{L^2}+\frac{d-2}{r^2}+\frac{2\kappa^2 p}{d-1}
\right\}
+\frac{i\omega e^{\chi/2}(p+\rho)}{w f^{3/2}}
 \right)\delta u^r(r)
 \nonumber \\
  & - \left( {\frac{d r}{2L^2fw}+\frac{\kappa^2 r p}{(d-1)fw}}
  +  \frac{r\kappa^2(2p+\rho)}{(d-1)f}+\frac{(d-2)(1-f)(1+w)}{2rf w}
+\frac{d r}{2L^2f}
\right)\delta\rho (r) \\
\delta {u^r}'(r) = & -
\left(
\frac{d-1}{r}+\frac{d-2}{2rw}-\frac{r}{2fw}\left\{ 
\frac{d}{L^2}+\frac{d-2}{r^2}+\frac{2\kappa^2 p}{d-1}
\right\}\right)\delta u^r(r)
+\frac{i\omega e^{\chi/2}}{\sqrt{f} (p+\rho)}\delta\rho(r)\ .\label{eq:genpert}
\end{align}

Compatibility with either the microcanonical or canonical ensemble imposes boundary conditions at $r \to \infty$.  General solutions in this limit have the boundary behavior 
\be
	\delta u^r \sim a/r^{d-1-1/w},~\delta\rho \sim b/ r^{\frac{1+w}{w}}\ .
\ee
In the microcanonical ensemble, where energy is conserved, we need to require the energy flux through the boundary to vanish.  The energy flux is
\be
\int_t^{t+\Delta t}dt  \int d\Omega_{d-1} T_{ij} (\partial_t)^i n^j \propto a\ ,\label{eq:boundaryflux}
\ee
where $n^j$ is a outward unit vector orthogonal to the $(r = {\rm constant})$ surface, so fixed energy imposes the condition $a=0$. In the canonical ensemble the boundary temperature has to be conserved. Given the expression (\ref{eq:w_temp_density}), we see that fixed temperature requires $b = 0$. Since $a$ is no longer fixed, Eq. (\ref{eq:boundaryflux})\ shows that energy can enter and leave the boundary at infinity.  We can think of this as exchanging energy with a heat bath coupled to the boundary that maintains thermal equilibrium.\footnote{Stability of the canonical ensemble is often found by studying negative eigenvalues of the wave operator in the Euclidean geometry for the finite-temperature solution.  This would require a Lagrangian description of the fluid. This will be studied in \cite{mmr_WIP}.}

\subsection{Stars in $AdS_3$} \label{subsec:AdS3}

For self-gravitating fluids in $AdS_3$, 
Einstein's equations can be solved exactly in the case of linear and polytropic equations of state \cite{Garcia:2002rn}.~\footnote{A derivation of these solutions from a 4d Randall-Sundrum construction can be found in \cite{Vaganov:2007at}.} 
We will make use of a different metric ansatz from (\ref{ansatz}).  In the  case $p=w\rho$ with  $0<w<1$, the solution is:

%
\begin{eqnarray} \label{3dmet}
ds^2 & = &-\frac{r^2}{L^2}dt^2+G L^2d\phi^2+\frac{dr^2}{G}  \nonumber \\
G & =& \frac{r^2}{L^2} - C - \frac{2\kappa^2w^2\mu}{(1-w)}r^\frac{w-1}{w}  \nonumber\\
\rho & = & \mu/r^\frac{1+w}{w}
\end{eqnarray}
We will assume that the energy density is positive, so that $\mu > 0$.  In this case, one can show that $G' > 0$ for all $r > 0$.  
Thus, all zeros of $G$ on the positive real axis are simple zeros.  Assume that the largest such zero is at $r = r_0$; we can write 
\be
	C = \frac{r_0^2}{L^2} - \frac{2\kappa^2w^2\mu}{(1-w)}r_0^\frac{w-1}{w}
\ee
There is a potential coordinate singularity at $r = r_0$, corresponding to the origin of polar coordinates.  Smoothness requires 
$r_{0} \leq r < \infty$ and 
\be
\mu = \frac{(L-r_0)r_0^{1/w}}{L^2\kappa^2 w}\ .
\ee
The local density at the core is therefore
\be
	\rho_0 = \frac{L-r_0}{L^2\kappa^2 w r_0}\ .
\ee
Positivity of the energy density requires $r_0 < L$. Since $\partial_{r_0} C > 0$, $C$ will increase monotonically with $r_0$, and as $r_0 \to 0$, $C \to - \infty$.  Thus we can demand $0 \leq r_0 \leq L$; in this range there is a one-parameter family of space times labeled by $r_0$.  At $r_0 = L$, we can define $r^2=\tilde r^2 +L^2$ to show that the metric is simply that of anti-de Sitter space in global coordinates.  For $r_0 = 0$, the metric is that of the ``zero mass'', zero temperature BTZ black hole.
 
We can now compute thermodynamic and stability properties of these solutions.  The energy is
\be
E =
\frac{\pi r_0 (2 w L - r_0(1+w))}{L^2\kappa^2(1-w)},\label{eq:ads3mass}
\ee
Global AdS space-time corresponds to $E = - \pi/\kappa^2$.  The energy takes its maximum value at  $r_{0,E_{max}} = wL/(1+w) < L$, at which value 
\be
	E_{max} = \frac{w^2}{1 - w^2} \frac{\pi}{\kappa^2} > 0
\ee
For smaller $r_0$, the mass decreases until $r_0 = 0, M= 0$.  There is also a solution with $M = 0$ at $r_{0,b} = \frac{2wL}{1+w} > r_{0,E_{max}}$.  For positive mass, there are two solutions.
The temperature and total entropy are
\begin{eqnarray}
	T & = & \frac{(r_0 \sigma)^\frac{1}{1+w} (L-r_0)^\frac{w}{1+w}}{L\sigma (L^2\kappa^2 w)^\frac{w}{1+w}}\nonumber\\
	S & = &  \frac{2\pi(1+w)L(w r_0 \sigma)^\frac{w}{1+w} (L-r_0)^\frac{1}{1+w}}{(1-w) (L \kappa)^\frac{2}{1+w}}\ .\label{eq:ads3tempent}
\end{eqnarray}
The temperature has a maximum at $r_{0,T_{max}} =L/(1+w) > r_{0,E_{max}}$; this occurs at a lower energy than $E_{max}$.  Interestingly, at this value of $r_0$, $\p_{r_0} \mu = 0$, and $\p_{r_0}^2 G = 0$. In particular, in a Taylor expansion of $G$ about $r = r_0$, the quadratic term vanishes. We believe this is a new result.  There should be a physical reason for this apparent coincidence, since it happens for all values of $w$, but we have not found any.

\begin{figure}[thbp]
\begin{center}
\includegraphics[scale=.5]{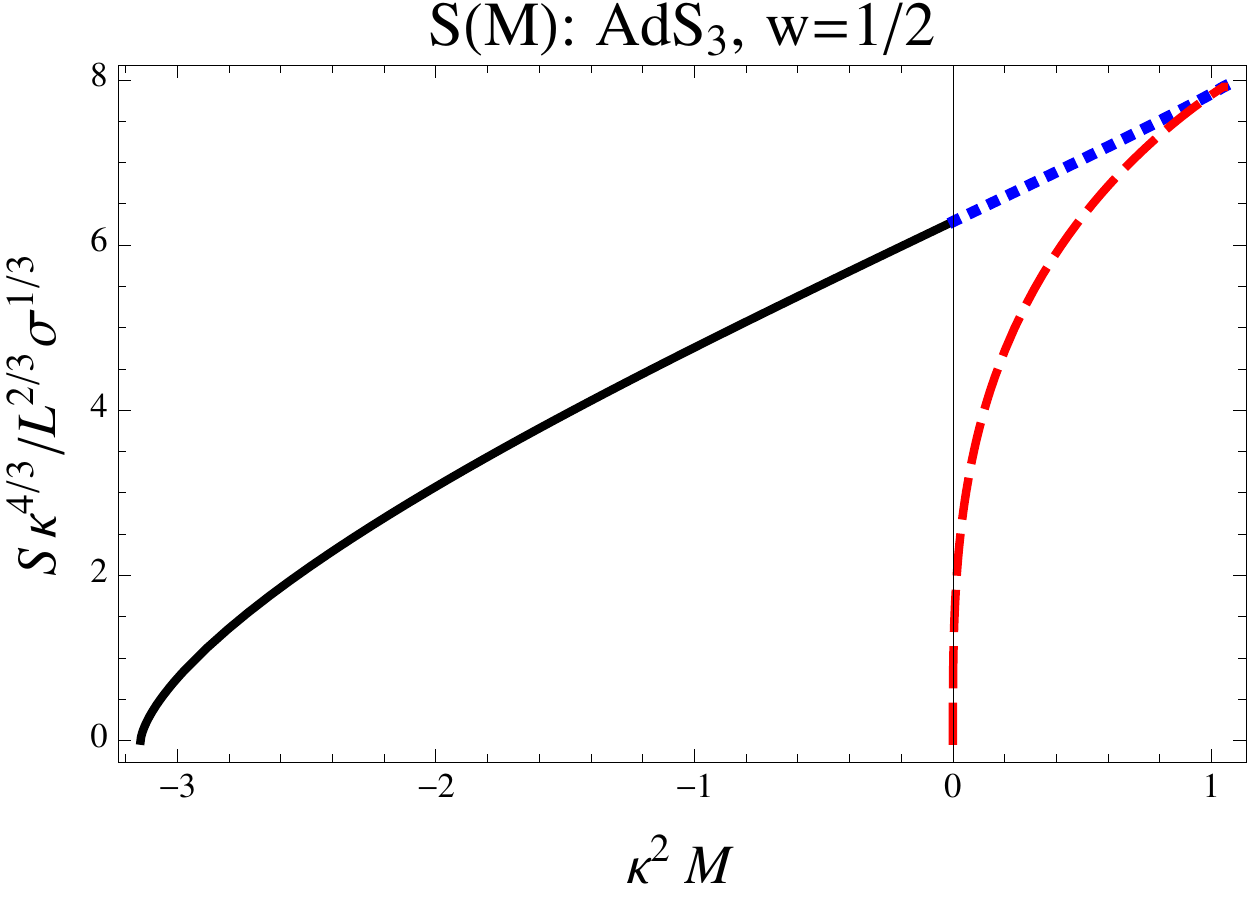}
\includegraphics[scale=.5]{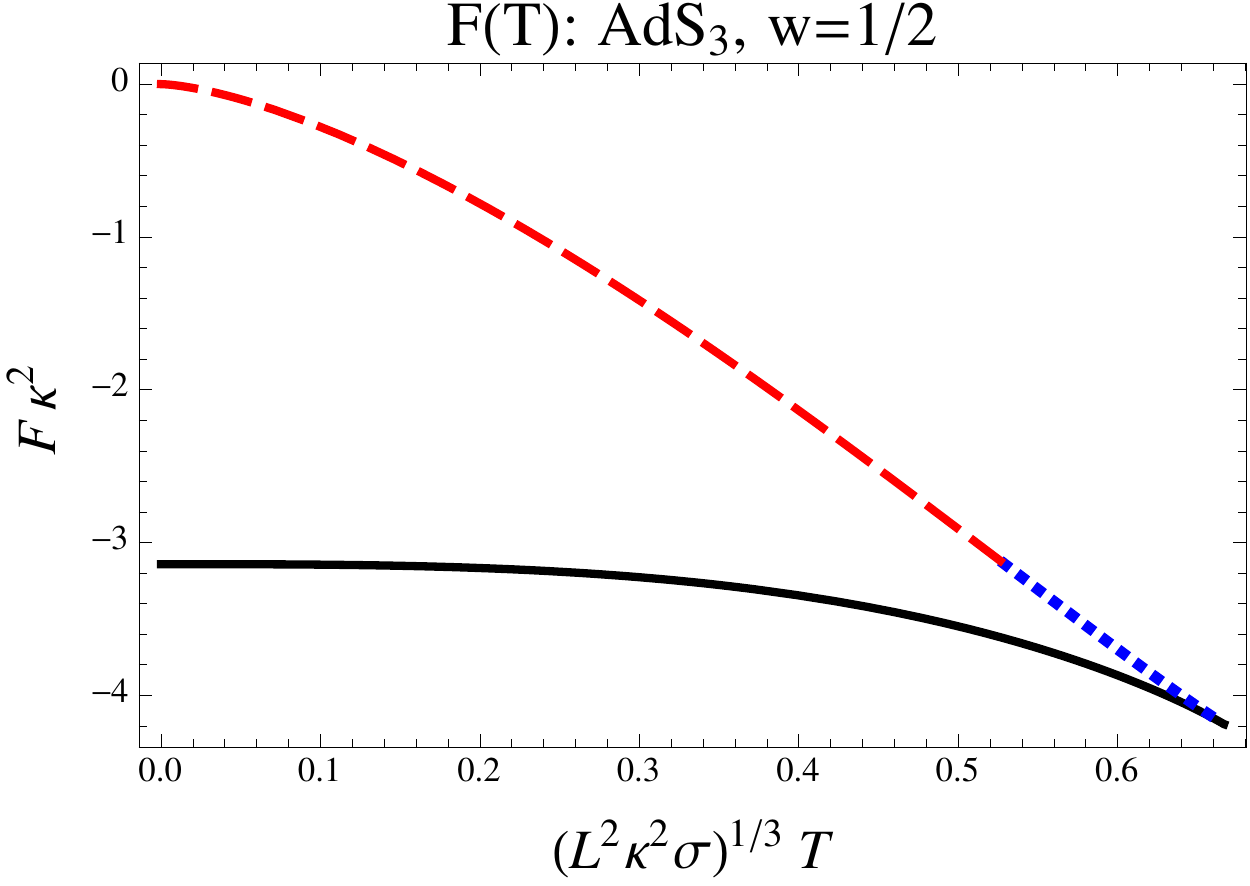}
\includegraphics[scale=.5]{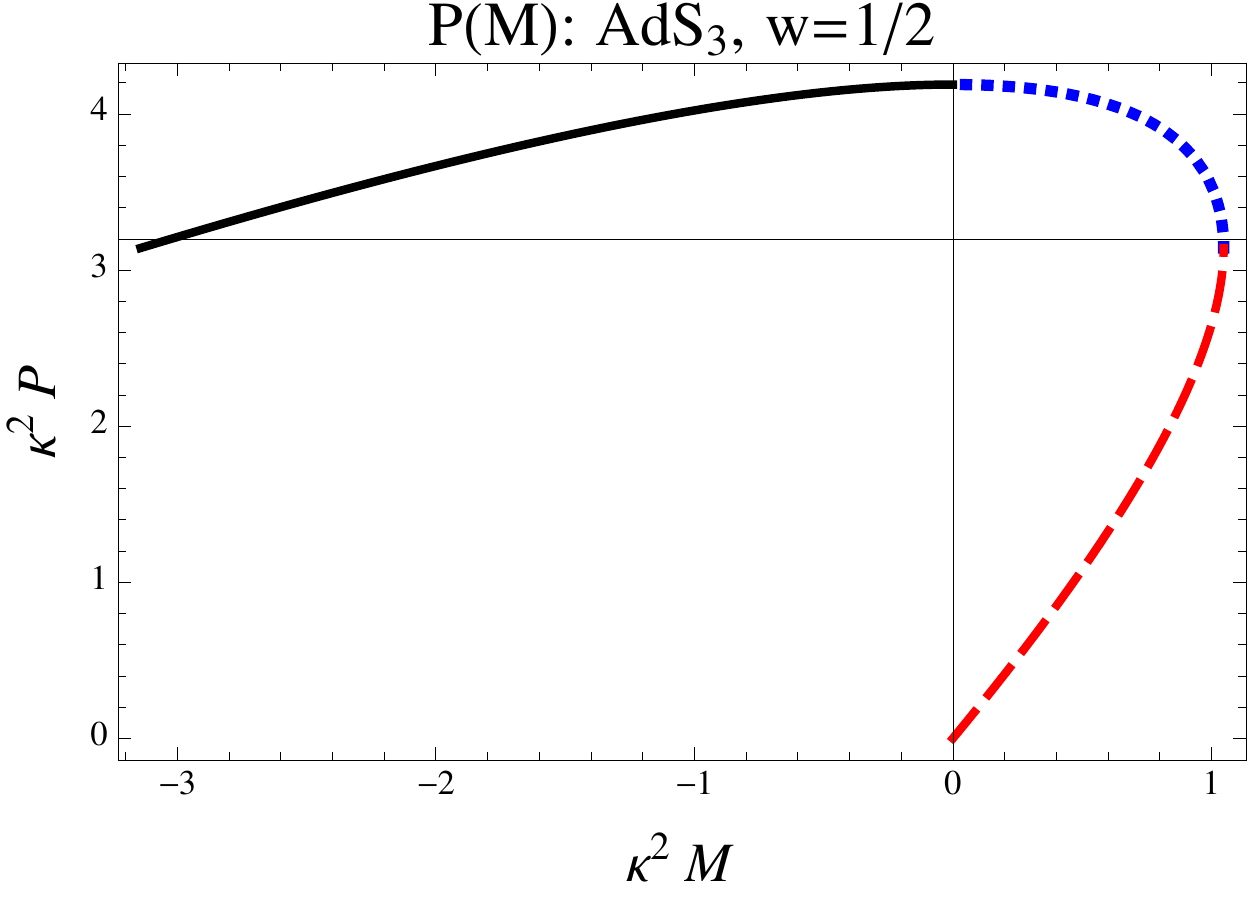}
\caption{The energy-entropy curve, temperature-free energy, and energy-pressure (boundary equation of state) curves for radiation stars in $AdS_3$. The solid black curve is both entropically and thermodynamically preferred, the blue dotted section is entropically preferred but thermodynamically disfavored, and the red dashed line is always disfavored.
\label{fig:ads3_mass_star}}
\end{center}
\end{figure}

The curves $S(E)$, $F(T)$, and $p(T)$ are plotted in \figref{fig:ads3_mass_star}.  For $r_{0,T_{max}} < r_0 < L$, $ - \frac{\pi}{\kappa^2} \leq E \leq 0$, and there is only a single solution at fixed $E$.  For $E \geq 0$, 
$0 \leq r_0, \leq r_{0,T_{max}}$ there are two branches of solutions, which meet at $r_0 = r_{0,E_{max}}$.  The branch $r_{0,E_{max}} \leq r_0 \leq r_{0,T_{max}}$ has higher entropy than the branch $0 \leq r_0 \leq r_{0,E_{max}}$.  On the other hand, as a function of temperature, there are always two solutions; one branch corresponds to $r_{0,T_{max}} \leq r_0 \leq L$, and has the lowest free energy.

Thus, for $E < 0$, $r_{0,T_{max}} < r_0 \leq L$, the solution is thermodynamically preferred amongst all spherically symmetric solutions without horizons, in both the canonical and microcanonical ensemble.  For $r_{0,E_{max}} < r_0 \leq r_{0,T_{max}}$, the solutions are preferred in the microcanonical ensemble but disfavored in the canonical ensemble.  Finally, for $0 \leq r_0 < r_{0,E_{max}}$, the solutions are disfavored in either ensemble.  

\begin{figure}[thbp]
\begin{center}
\includegraphics[scale=.5]{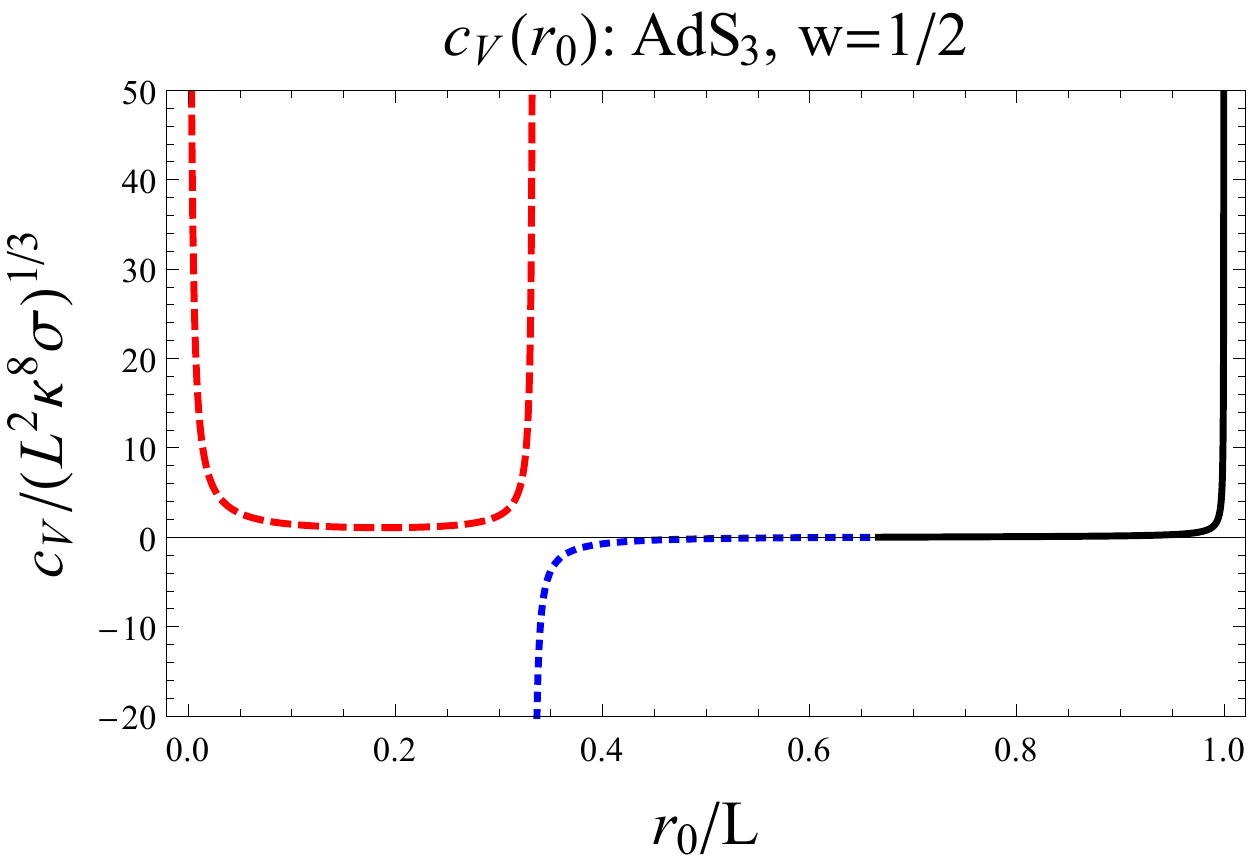}
\includegraphics[scale=.5]{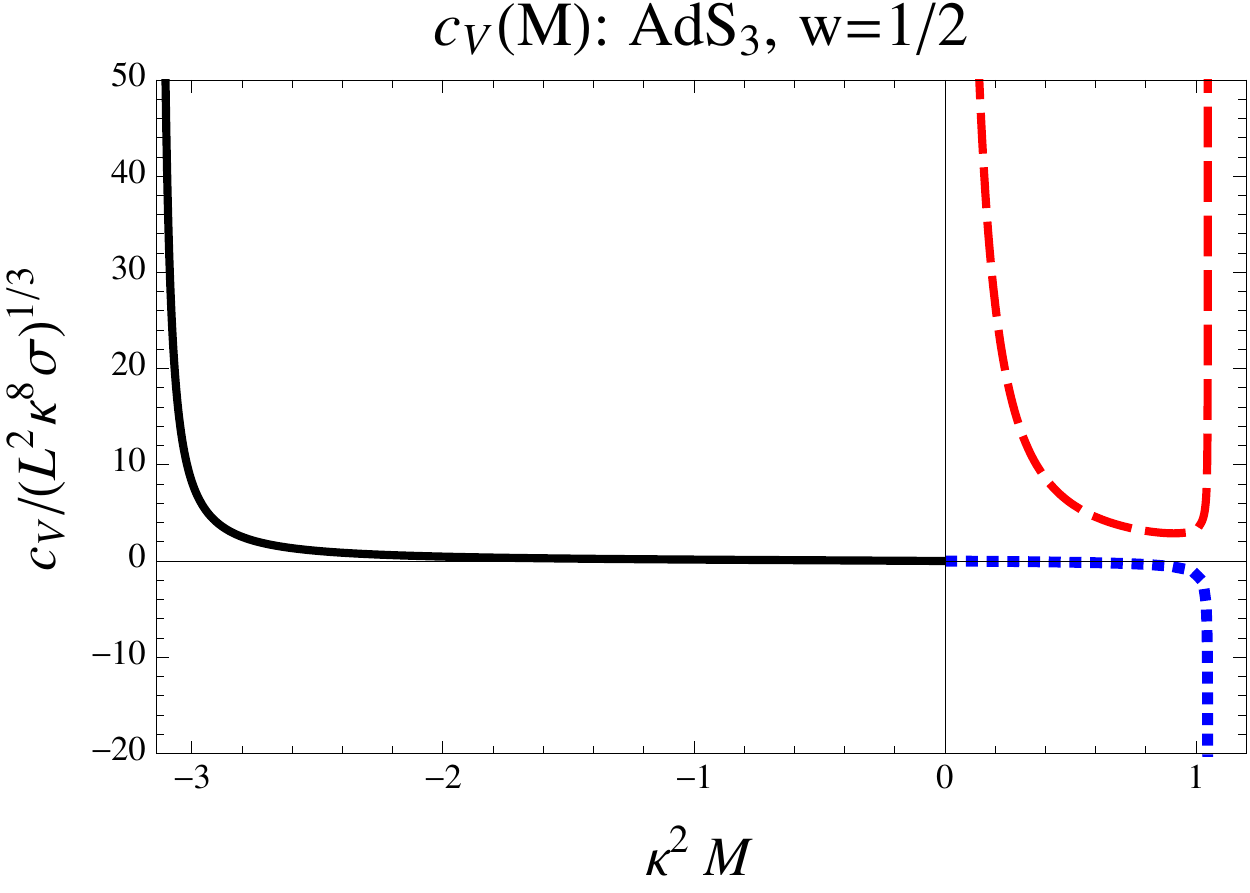}
\caption{Specific heat as a function of the parameter $r_0$ (left figure) and mass (right figure). The specific heat changes sign through a pole at the maximum temperature solution, and changes sign through a zero at the maximum mass point. As above, the solid black curve is both entropically and thermodynamically preferred, the blue dotted curve is entropically preferred but thermodynamically disfavored, and the red dashed line is canonically and microcanonically disfavored.
\label{fig:ads3_specific_heat}}
\end{center}
\end{figure}

We can also compute the microcanonical specific heat 
$c_{V} \equiv dE/dT = -(T^2 d^2S/dE^2)^{-1}$, which  is
\be
c_{V}=
\frac{
2\pi (w\sigma)^\frac{w}{1+w}(1+w)
}{
(1-w)L^\frac{1-w}{1+w}\kappa^\frac{2}{1+w}
}
\frac{
r_0^\frac{w}{1+w}(L-r_0)^\frac{1}{1+w}(L w - r_0 (1+w))
}{
L-(1+w)r_0
}
\ee
We have plotted this as a function of $r_0$ and as a function of $E$ in \figref{fig:ads3_specific_heat}. We can already see from the equation above that $c_V$ starts positive at $r_0 = L$ and becomes negative through a pole for $r_{0,E_{max}}< r_0 < r_{0,T_{max}}$, which is the regime at which the temperature begins to drop as the mass continues to increase, and in which the configuration is the preferred solution in the microcanonical ensemble but disfavored in the canonical ensemble.  For $r_0 < r_{0,E_{max}}$, $c_V$ becomes positive again through a zero; this is the regime in which the solution is disfavored in either ensemble.

\subsubsection{Dynamical stability}\label{ads5stab}

The metric ans\"atz \eqref{3dmet} is in a different gauge than \eqref{ansatz} but under a change of coordinates\footnote{The change of coordinates is going from a gauge where $g_{\phi\phi}=r^2$ to one where $g_{tt}=-r^2/L^2$, which is just a redefinition of the radial coordinate.} (\ref{eq:genpert}) gives:
\begin{eqnarray}
\delta {u^r}' & = & \frac{1-w+2L^2\kappa^2 w \rho}{w r (1+L^2\kappa^2 w \rho)}\delta u^r + \frac{i\omega L}{r \rho (1+w)}\delta\rho\nonumber\\
\delta\rho' &=& i \left( \frac{(1+w)\omega L\rho}{w r G} +\frac{2\kappa^2 L(1+w)^2\rho^2}{\omega r w (1+L^2\kappa^2 w \rho)}\right)\delta u^r-
\frac{(1+w)(1+2L^2\kappa^2 w \rho)}{ r w (1+L^2\kappa^2 w \rho)}\delta\rho\ .\label{eq:ads3perts}
\end{eqnarray}
As discussed in Section \ref{subsubsec:mechstab}, fixed mass requires that the leading term in $\delta u^r =a r^{1/w-1}+\ldots$ vanish, while fixed temperature requires that the leading term in $\delta\rho = b/r^{1+1/w}+\ldots$ vanish. For a given set of boundary conditions, eq. (\ref{eq:ads3perts}) can be thought of as eigenvalue equations for $\omega$.

\begin{figure}[thbp]
\begin{center}
\includegraphics[scale=.5]{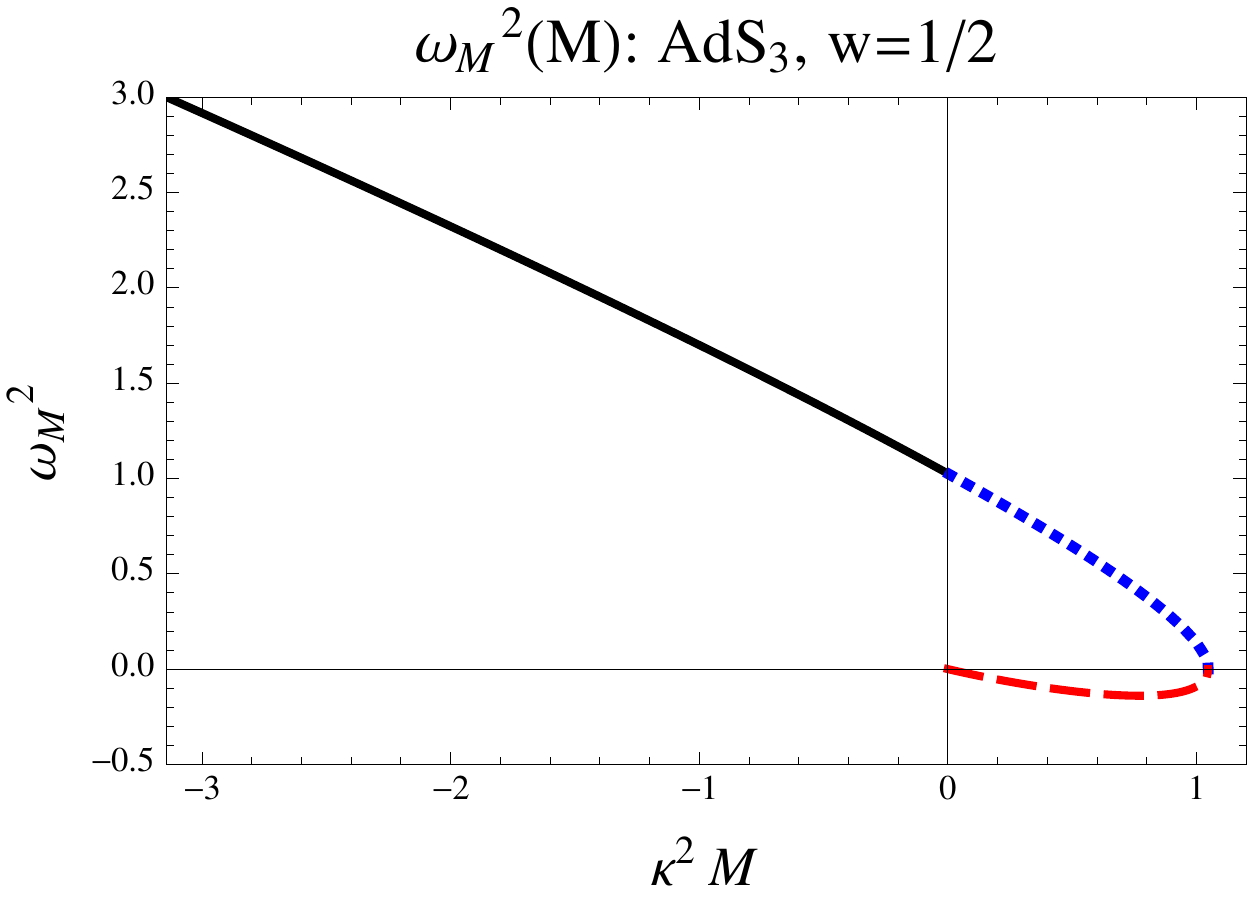}
\includegraphics[scale=.5]{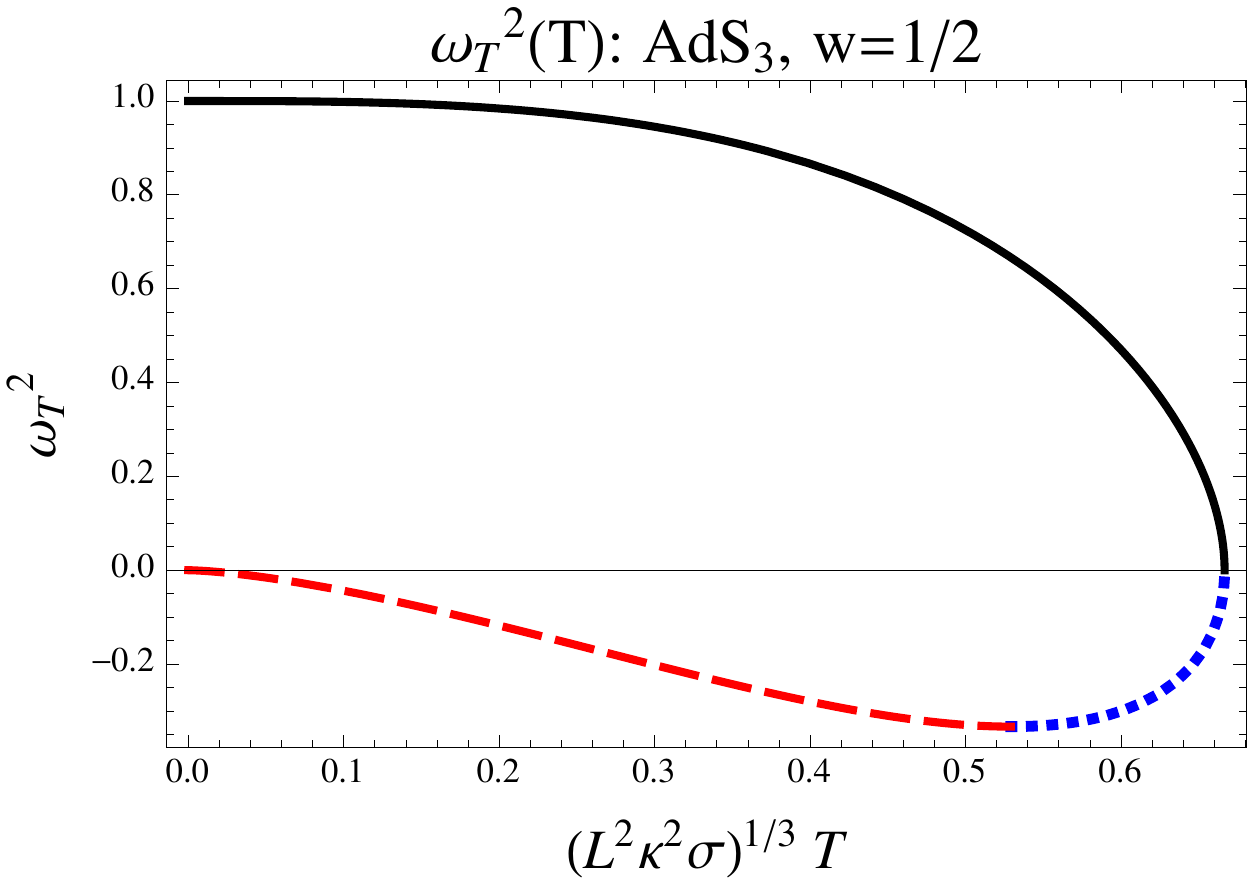}
\caption{Frequency of the lowest normal mode in the microcanonical (left figure) and  canonical (right figure) ensemble, for a self-gravitating gas of radiation in an asymptotically $AdS_3$ background. Recall following figure \ref{fig:ads3_mass_star} that the black solid line is thermodynamically and entropically preferred, the blue dotted line is entropically preferred but thermodynamically disfavored, and the red dashed line is canonically and microcanonically disfavored.   
\label{fig:ads3_dynamic_instability}}
\end{center}
\end{figure}

We find instabilities for both $a=0$ and $b = 0$ at sufficiently high core density.  In the case $a = 0$, corresponding to perturbations in the microcanonical ensemble, let $\omega_E$ denote the lowest frequency consistent with Eq. (\ref{eq:ads3perts}).  An instability sets in when $\omega_E^2$ becomes negative. As we can see in \figref{fig:ads3_dynamic_instability}, this occurs precisely at the solution with maximum mass, $E_{max} \sim {\cal O}(1/\kappa^2)$.  This is consistent with the extension to asymptotically anti-de Sitter spacetimes \cite{Roupas:2013nt}\ of the flat space results of \cite{Green:2013ica}: the appearance of a turning point in $E$, as a function of some parameter labeling the solutions (such as the core density), indicates the onset of an instability. Note also that this instability sets in precisely when the solution ceases to be a maximum of the entropy $S$, even ignoring black hole solutions, as we can see from \figref{fig:ads3_mass_star}. This is reminiscent of the results of \cite{Sorkin:1981wd}\ for asymptotically flat spacetimes.  In that work, mechanical stability was correlated with the entropy being a {\it local} maximum with respect to local perturbations. We have found that at the onset of instability, a higher entropy solution exists at fixed mass.

In the case $b = 0$, let $\omega_T$ the lowest frequency consistent with Eq. (\ref{eq:ads3perts}).  As can be seen in Fig. \figref{fig:ads3_dynamic_instability}, the instability appears when the {\it temperature} reaches a maximum as a function of the core density, while the mass is still growing.  This corresponds to the onset of negative specific heat.  This occurs when $\sigma T_{max}^3 \sim {\cal O}(1/(L^2\kappa^2))$.  Inspection of \figref{fig:ads3_specific_heat} and \figref{fig:ads3_dynamic_instability}\ show that at a still higher value of the core density, the specific heat becomes positive again, but the instability persists. As can be seen from \figref{fig:ads3_mass_star}, the solution ceases to minimize the free energy even ignoring black hole solutions at the same temperature; the fact that the instability remains indicates that the solution is not even a local minimum of the free energy.

\subsection{Stars in $AdS_5$} \label{subsec:AdS5}

Next we consider radiation in asymptotically $AdS_5$ backgrounds ($d=4$ and $p=\rho/4$), relevant for the physics of four-dimensional large-N, strongly coupled gauge theories. In particular, such solutions describe the ``confined'' phase of these gauge theories, which dominate the canonical thermodynamics below the Hawking-Page/deconfinement transition, and are metastable above it \cite{Witten:1998qj,Witten:1998zw}.   Unfortunately exact solutions are not available, so we must resort to approximations and numerical solutions.

We can warm up by computing the entropy and free energy functions, perturbatively in the asymptotic value of the density. Writing $\rho=\mu/r^5+\ldots$, we find
\begin{eqnarray}
E & = & \frac{4\pi^2}{3L}\mu+\frac{\pi^2\kappa^2}{16L^4}\mu^2+\frac{7\pi^2\kappa^4(5040 \log(2)-3487)}{2916L^7}\mu^3+\ldots\nonumber\\
S & = & \sigma^{1/5}\mu^{4/5}\left(\frac{5}{3} +\frac{5\pi^2\kappa^2}{72L^3}\mu+\frac{5\pi^2\kappa^4(5040\log(2)-3487}{1944L^6}\mu^{2} +\ldots\right)
\end{eqnarray}
Note that the expansion is in powers of $\kappa^2 \mu/L^3 \sim \mu/N^2$ for the dual to strongly-coupled, large-N ${\cal N}=4$ super-Yang-Mills theory \cite{Maldacena:1997re}.  We can combine these results to get the leading ${\cal O}(\kappa^2) ={\cal O}(1/N^2)$ corrections to the energy-entropy scaling for a self-gravitating gas of particles.
\be
S=\frac{5\pi^{2/5}\sigma^{1/5}}{2^{8/5}3^{1/5}}E^{4/5}\left(1+\frac{\kappa^2 E}{320 \pi^2L^2} +\ldots\right)
\ee
as well as the the temperature-free energy scaling,
\be
F=-\sigma\frac{L^4\pi^2}{3}T^5\left(1+ \sigma L^2\kappa^2T^5/48+\ldots\right)
\ee
Note that the corrections to the leading $S \sim E^{4/5}$ occur in a power series in $\kappa^2 E/L^2 \sim (EL)/N^2$, and the corrections to the leading $F(T) \sim \sigma L^4 T^5$ are a power series in $\sigma L^2 \kappa^2 T^5 \sim \sigma (LT)^5/N^2$. Following the discussion of instabilities in $AdS_3$, we might expect instabilities to set in when the entropy and free energy have non analytic behavior, where we would expect the above power series to break down; this should occur at $EL \sim L^3/\kappa^2 \sim N^2$ and $\sigma (LT)^5 \sim N^2$. We will show below that this scaling is correct.

\begin{figure}[thbp]
\begin{center}
\includegraphics[scale=.25]{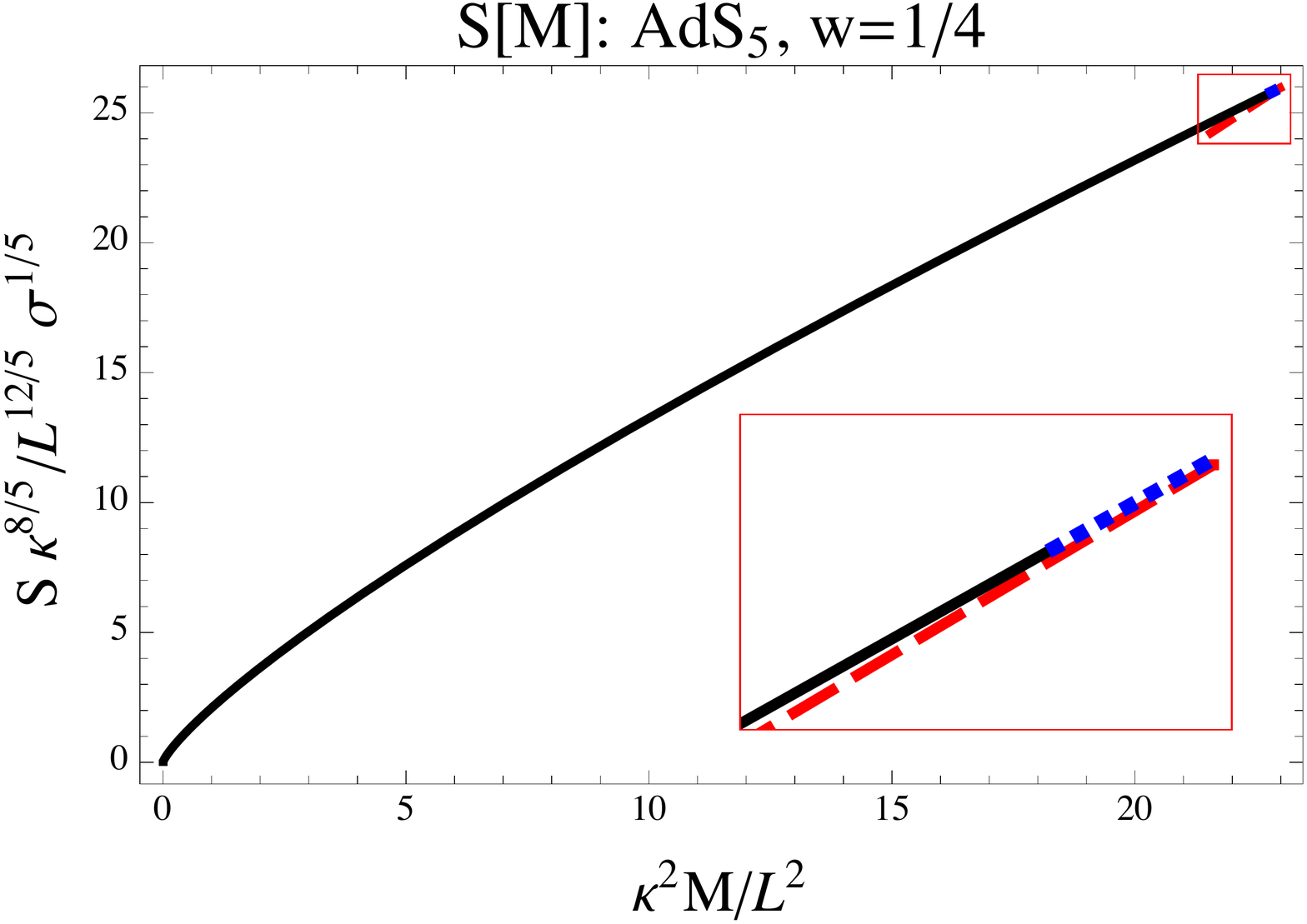}
\includegraphics[scale=.25]{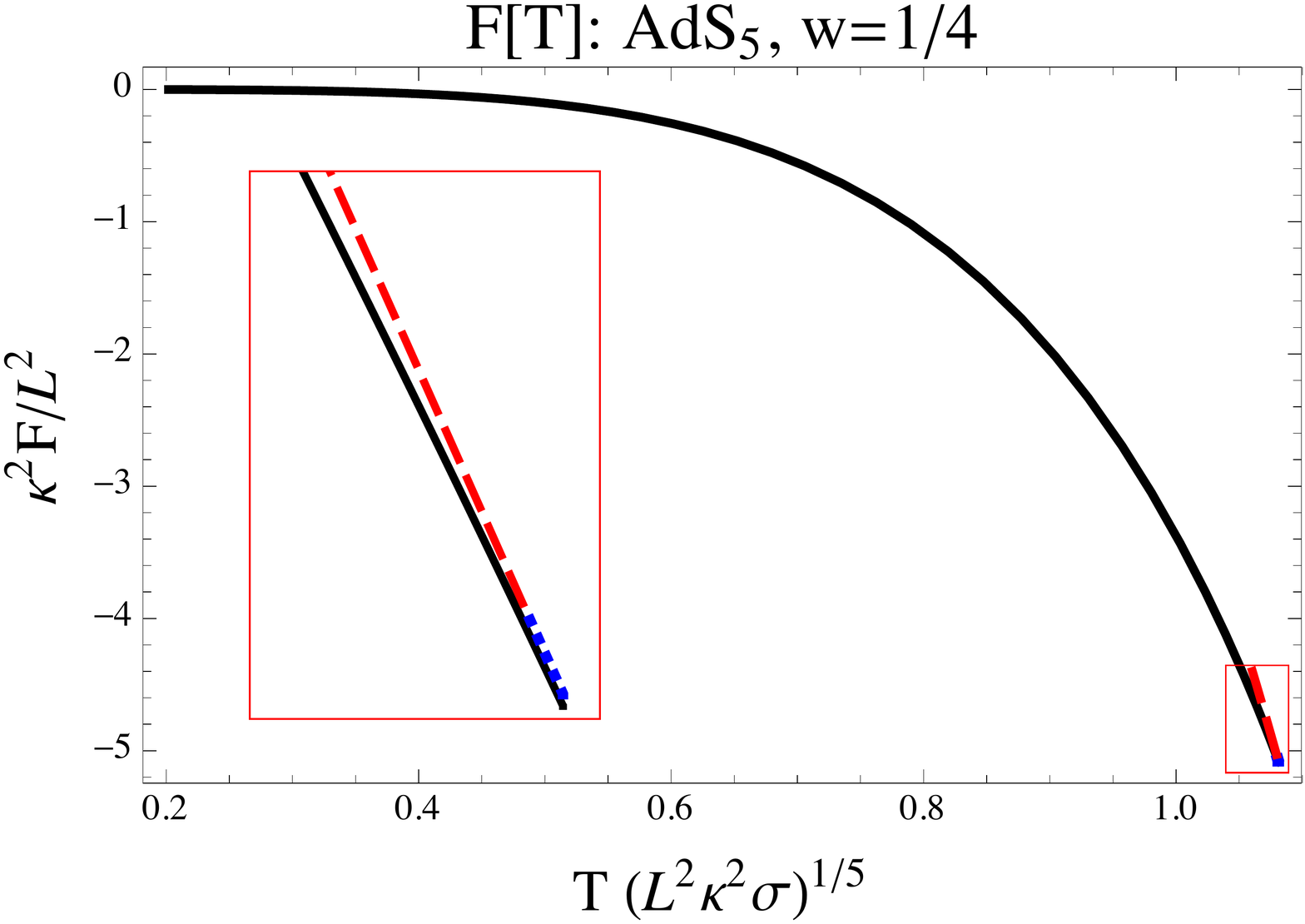}
\includegraphics[scale=.5]{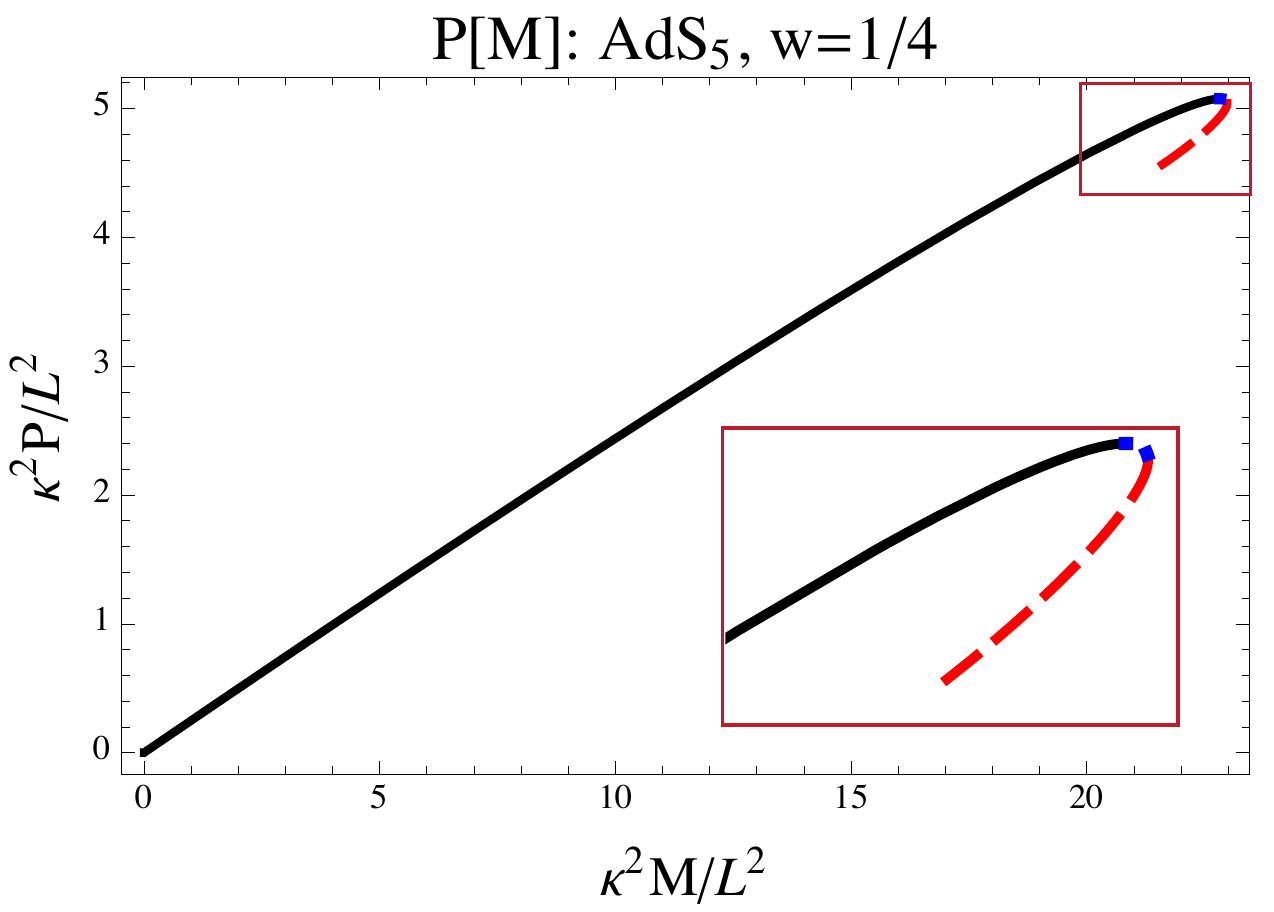}
\caption{
The energy-entropy curve, temperature-free energy, and total pressure-energy (boundary equation of state) curves for stars in $AdS_5$. The solid black curve is both entropically and thermodynamically preferred, the blue dotted section is entropically preferred but thermodynamically disfavored, and the red dashed line is always disfavored. The structure is the same as for the $AdS_3$ case, Fig. \ref{fig:ads3_mass_star}, though we have to zoom in because the dotted blue section is much smaller.
\label{fig:ads5_mass_star}}
\end{center}
\end{figure}

We can also construct the full nonlinear solutions numerically.  The results for entropy $S(E)$, free energy $F(T)$, and pressure $P(E)$ are shown in
\ref{fig:ads5_mass_star}.  The structure is indeed much the same as the $AdS_3$ case. There 
are two branches as a function of either mass or temperature; the branches join at a maximum entropy and energy for $S(E)$ and at the minimum free energy and maximum temperature 
for $F(T)$. Again, the maximum entropy/maximum energy configuration occurs at a higher core density than the minimum free energy/maximum temperature configuration.

\begin{figure}[thbp]
\begin{center}
\includegraphics[scale=.5]{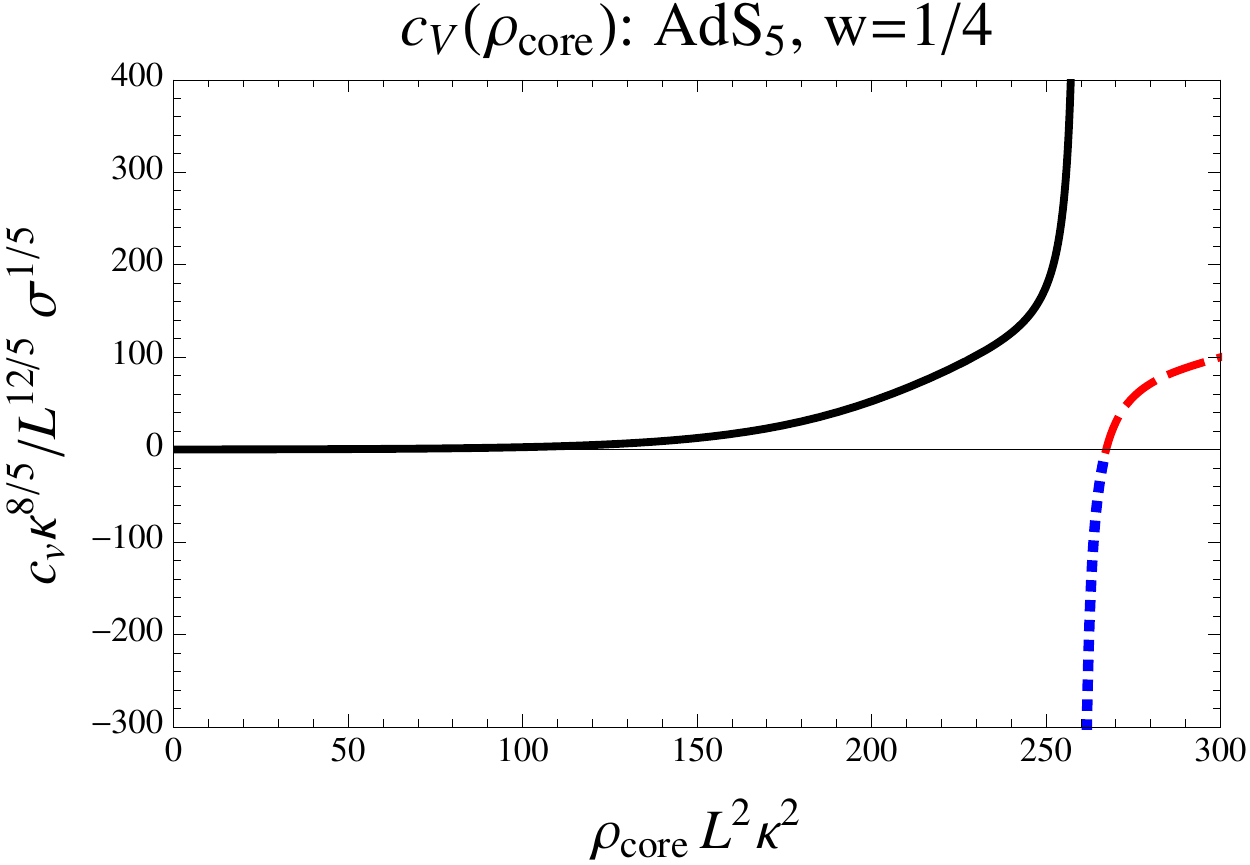}
\includegraphics[scale=.5]{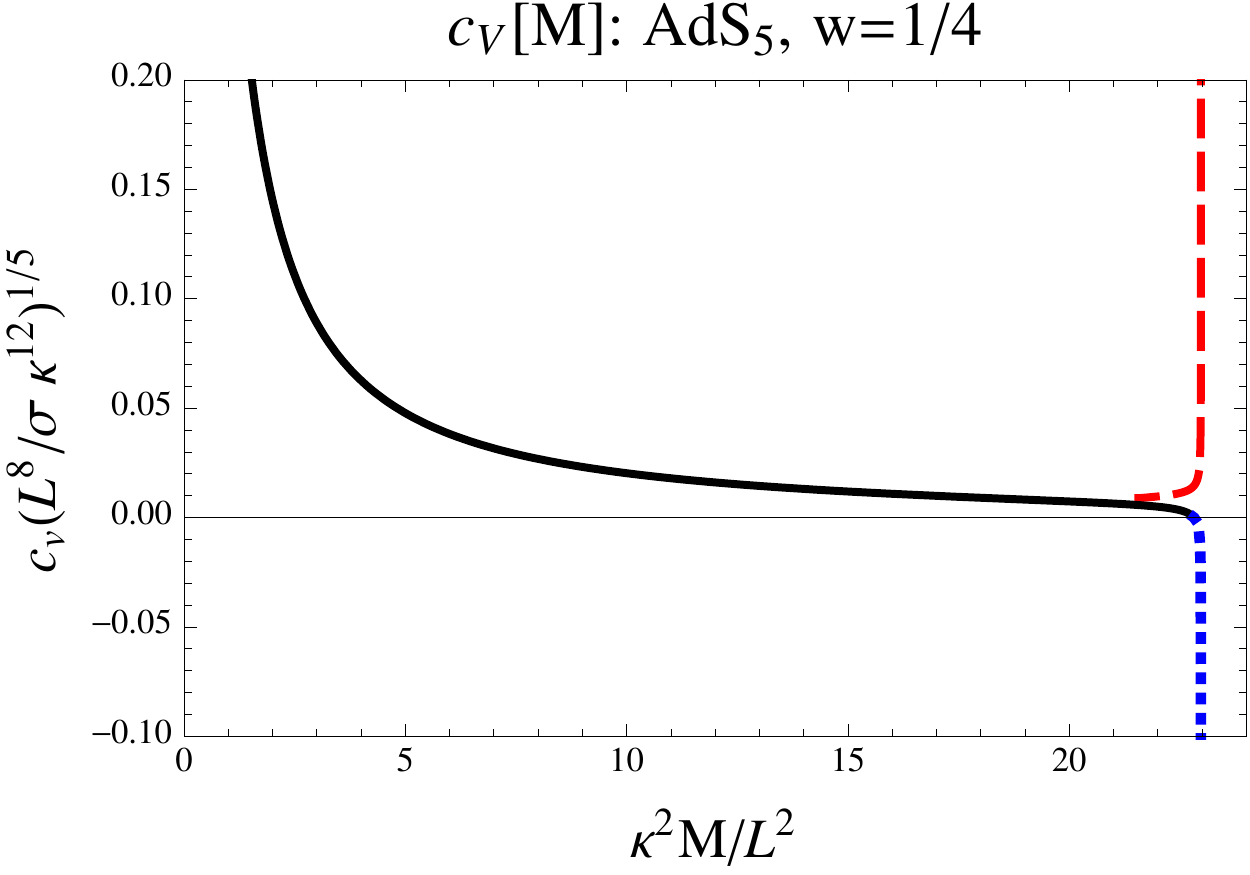}
\caption{
The microcanonical specific heat $c_V$ as a function of the core density (left figure) and energy (right figure).
\label{fig:ads5_specific_heat}}
\end{center}
\end{figure}

\figref{fig:ads5_specific_heat} shows the microcanonical specific heat as a function of core density and energy.  The specific heat also turns negative at core density at which the free energy ceases to be a minimum among static solutions with self-gravitating radiation; it becomes positive again at the core density at which the entropy ceases to be a maximum.

\subsubsection{Dynamical stability}

Following the discussion of Section \ref{subsubsec:mechstab} and \ref{ads5stab} we can determine the dynamical stability of these solutions. As before, we denote the lowest frequency squared for fixed-energy perturbations as $\omega_E^2$ and for fixed-temperature perturbations as $\omega_T^2$; instabilities set in when these become negative.  The numerical results are plotted in \figref{fig:ads5_dynamic_instability} below.  Our numerics indicate that the results are qualitatively the same as for $AdS_3$.  $\omega_E^2$ becomes negative, signaling an instability in the microcanonical ensemble, at the core density at which the energy and entropy are both at their maximum in this family of solutions.  $\omega_T^2$ becomes negative, signaling an instability in the canonical ensemble, at the core density at which the specific heat becomes negative, the free energy is at the minimum in this family of solutions, and the temperature is the maximum in this family of solutions.

\begin{figure}[thbp]
\begin{center}
\includegraphics[scale=.5]{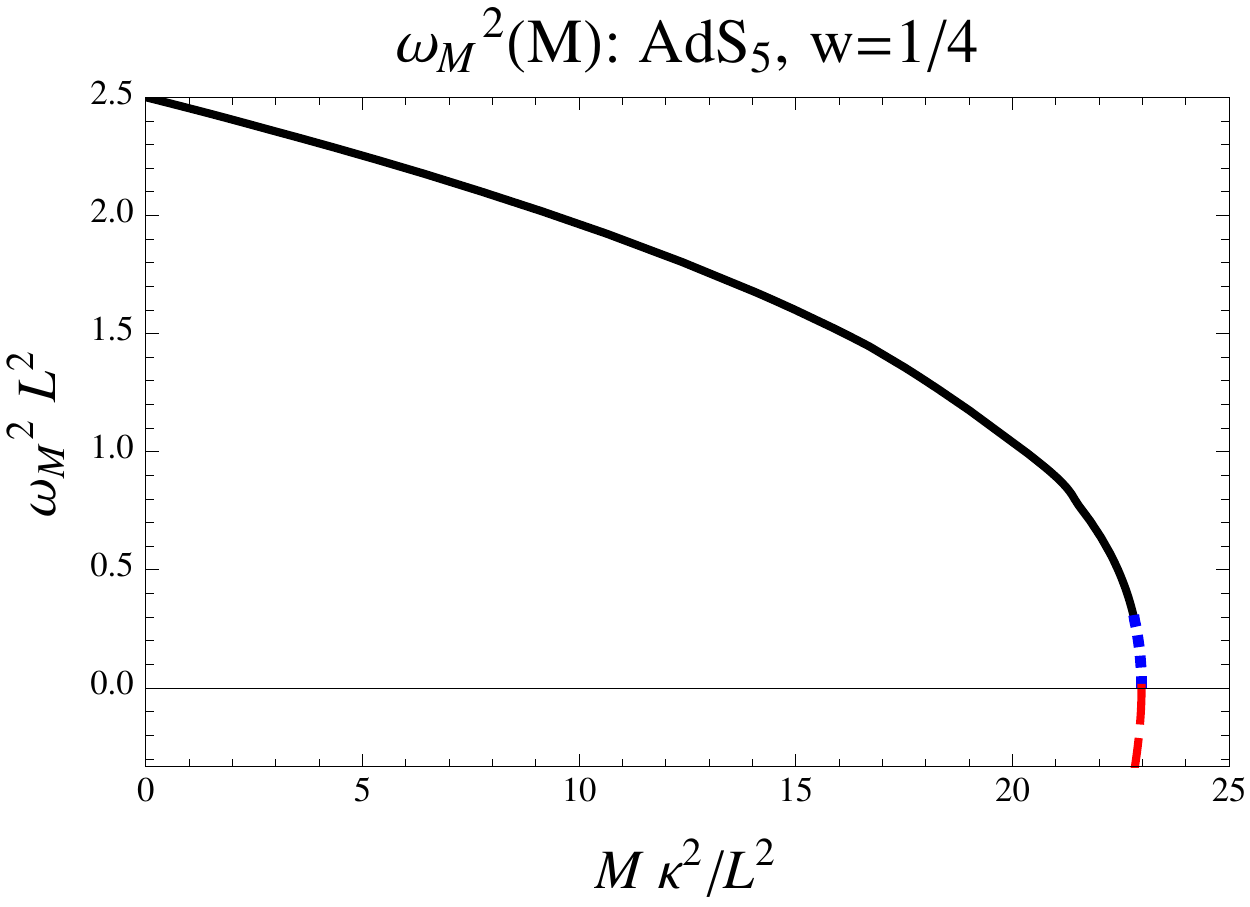}
\includegraphics[scale=.5]{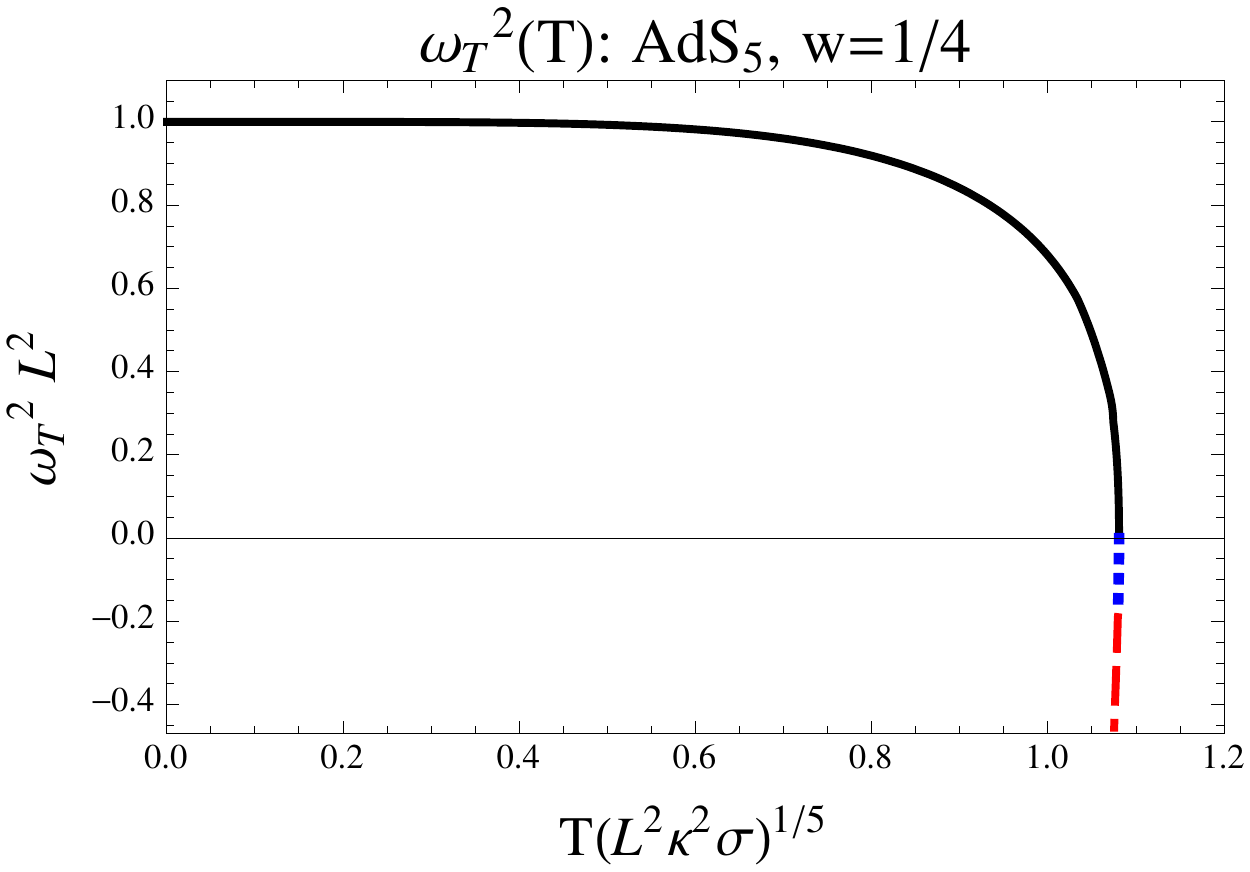}
\caption{Frequency of the lowest normal mode in the microcanonical (left) and canonical (right) ensemble for $AdS_5$ radiation stars. As before the black solid line is thermodynamically and entropically preferred, the blue dotted line is entropically preferred but thermodynamically disfavored, and the red dashed line is canonically and micro canonically disfavored.   The structure is the same as for the $AdS_3$ case Fig. \ref{fig:ads3_dynamic_instability}, though we have to zoom in as the dotted blue section is much smaller.
\label{fig:ads5_dynamic_instability}}
\end{center}
\end{figure}

\subsection{Stars in $AdS_{p} \times S_{q}$}

In the Freund-Rubin type compactifications that appear in gauge-gravity duality, the negative curvature of the $AdS_p$ factor is balanced by positive curvature in an $S_q$ factor plus flux.  In several well-known examples the radius of curvature of both factors is the same.  The phenomena we are interested in -- the temperature and mass at which various solutions to self-gravitating radiation become unstable -- occur at energies and temperatures much higher than this radius of curvature, by a factor of order $L^{p-2}/\kappa^2$.  Thus, we should check that the inclusion of this extra energy does not significantly change the solutions we have constructed above. We will focus on the case of $AdS_5\times S^5$ and $AdS_3\times S^5$.

The potential complication is that the backreaction of the thermal gas can interfere with the stabilization mechanism for the $S_q$ factor.  Thus we will specifically include the radius of the $S_q$ in the equations of motion that we integrate. However, assuming $S^q$ symmetry we know there is a Kaluza-Klein truncation to the $AdS_p$ system coupled to a massive dilaton as well as to the fluid (though the coupling to the fluid is modified by the dilaton), and we don't expect a massive field to change the phase structure much.

\subsubsection{$AdS_5\times S_5$ stars}

We begin with a truncation of IIB supergravity to the metric  and the self-dual five-form flux; this has the Freund-Rubin compactification to $AdS_5\times S_5$ as its vacuum solution. The equations of motion are \cite{Schwarz:1983qr,Howe:1983sra,Horowitz:1991cd}
\be
R_{ab}=\frac{1}{6}F_{ac_1c_2c_3c_4}F_b{}^{c_1c_2c_3c_4},~F=* F,~dF=0.\label{eq:IIB_ads_eq}
\ee
Global $AdS_5\times S_5$ is a solution with $F =L^4(1+*)\epsilon_5$,~$L_{AdS_5}=L_{S_5}=L$, where $\epsilon_5$ is the volume form on a unit five-sphere (that is, $\int_{S_5}\epsilon_5=\pi^3$). 
When we couple the theory to the fluid (\ref{fluidtensor}), eqn. (\ref{eq:IIB_ads_eq})
becomes
\be
R_{ab}=\frac{1}{6}F_{c_1c_2c_3c_4a}F_b{}^{c_1c_2c_3c_4}+\kappa^2\left(T_{ab}-T^{c}{}_cg_{ab}/8\right).
\label{eqn:EinstFlux}
\ee
The metric ans\"atz is now (\ref{ansatz}) plus an additional $5-$sphere whose
size depends on the AdS radial coordinate $r$:
\be
ds^2 = ds_5^2 + L^2e^{2 C(r)} d\Omega_5^2.\label{eq:ads5s5metric}
\ee
After some rearrangements, the equations of motion become:
\be
\frac{(3+5rC')p_{,\rho}}{p+\rho}\rho'-5C'(3+2 r C')-\frac{3(f-1)}{fr}+\frac{\kappa^2 r p}{f}+\frac{2(5e^{8C}-2)r}{e^{10 C}L^2f}=0,\label{eq:FR_rho_eom}
\ee
\be
(3+5 r C')f'+\frac{2(f-1)(3+10 r C')}{r}+30 r f C'^2-\frac{40r^2 C'}{e^{10 C}L^2}
\nonumber\ee
\be
+\frac{4(5e^{8C}-8)r}{e^{10 C}L^2}-\frac{r\kappa^2 }{4}\left[5 r (p-\rho)C'-(5p+3\rho)\right]=0,\label{eq:FR_f_eom}
\ee
\be
(3+5 r C')\chi'+\frac{80r\sinh(4C)}{L^2fe^{6C}}-10\left(1-rC'+\frac{2}{f}+\frac{4r^2}{f e^{10C}L^2} \right)C'
\nonumber\ee
\be
-\frac{\kappa^2}{4f}\left[ 5r^2(p-\rho)c'-r(13p+3\rho) \right]=0,\label{eq:FR_chi_eom}
\ee
\be
r f C''+f C'+\frac{4r}{e^{10C}L^2}(1-e^{8C})+2C'\left(1+\frac{2r^2}{e^{10C}L^2} \right)-\frac{r\kappa^2}{8}(p-\rho)(1-rC')=0.\label{eq:FR_C_eom}
\ee
Note that $\kappa^2$ denotes the {\it ten}-dimensional Planck scale in these equations.  After using the equation of state to determine $p(\rho)$, we have three first order and one {\it second} order equations in four unknowns $\rho, \chi, f$ and $C$; this is because we have exhausted all of the constraints from diffeomorphism invariance.  Finally, since we will be interested in energies and temperatures well above the scale $1/L$, we will consider the equation of state $p=\rho/9$, corresponding to radiation in ten dimensions.

\begin{figure}[thbp]
\begin{center}
\includegraphics[scale=.5]{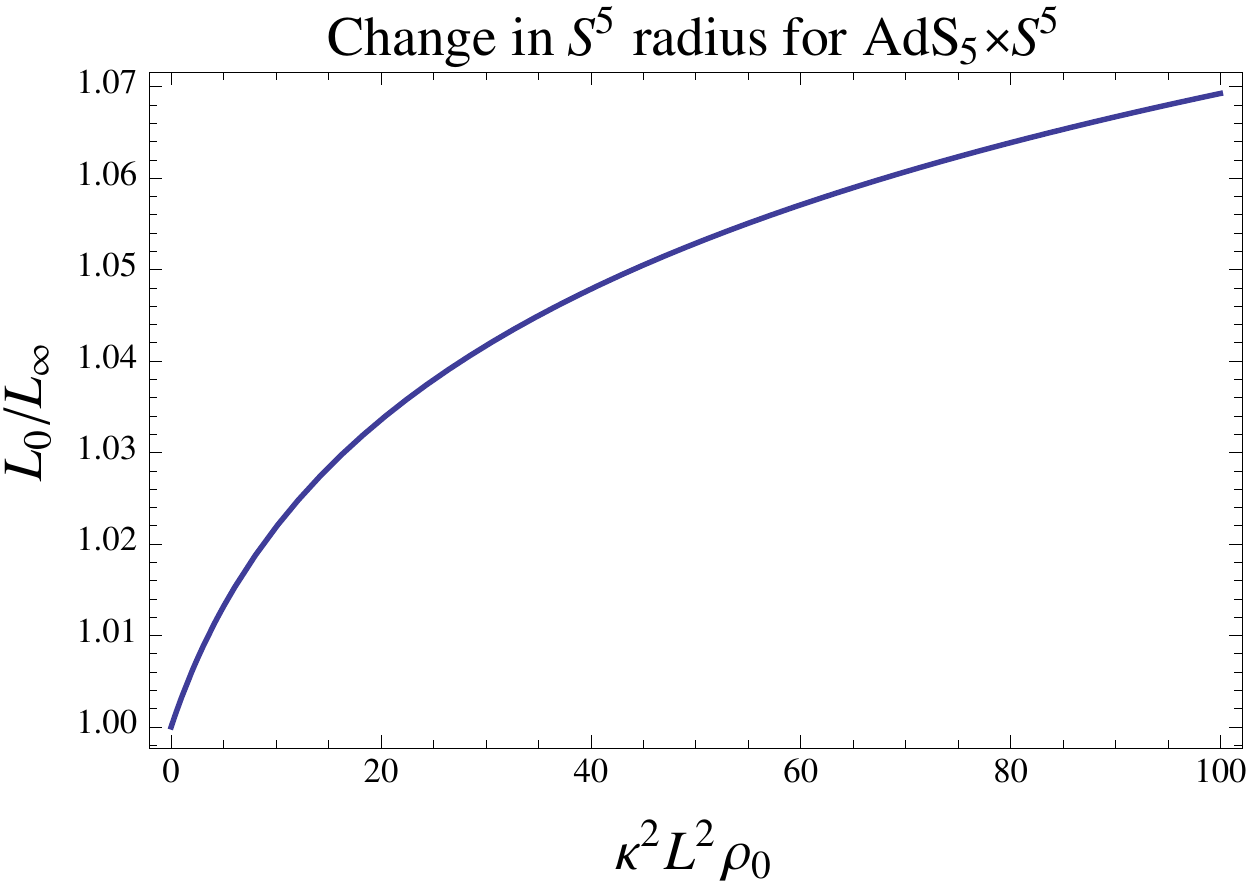}
\caption{
A plot demonstrating the small change in the $S_5$ size due to back reaction as a function of core density $\rho_0$. Note that the star becomes thermodynamically unstable at $L^2\kappa^2 \rho_0 \approx 10$ and microcanonically unstable at $L^2\kappa^2 \rho_0 \approx 25.$ It is clear that the inclusion of the five-sphere doesn't qualitatively change the structure of the solution.
\label{fig:ads5s5_radius_change}
}
\end{center}
\end{figure}

\begin{figure}[thbp]
\begin{center}
\includegraphics[scale=.5]{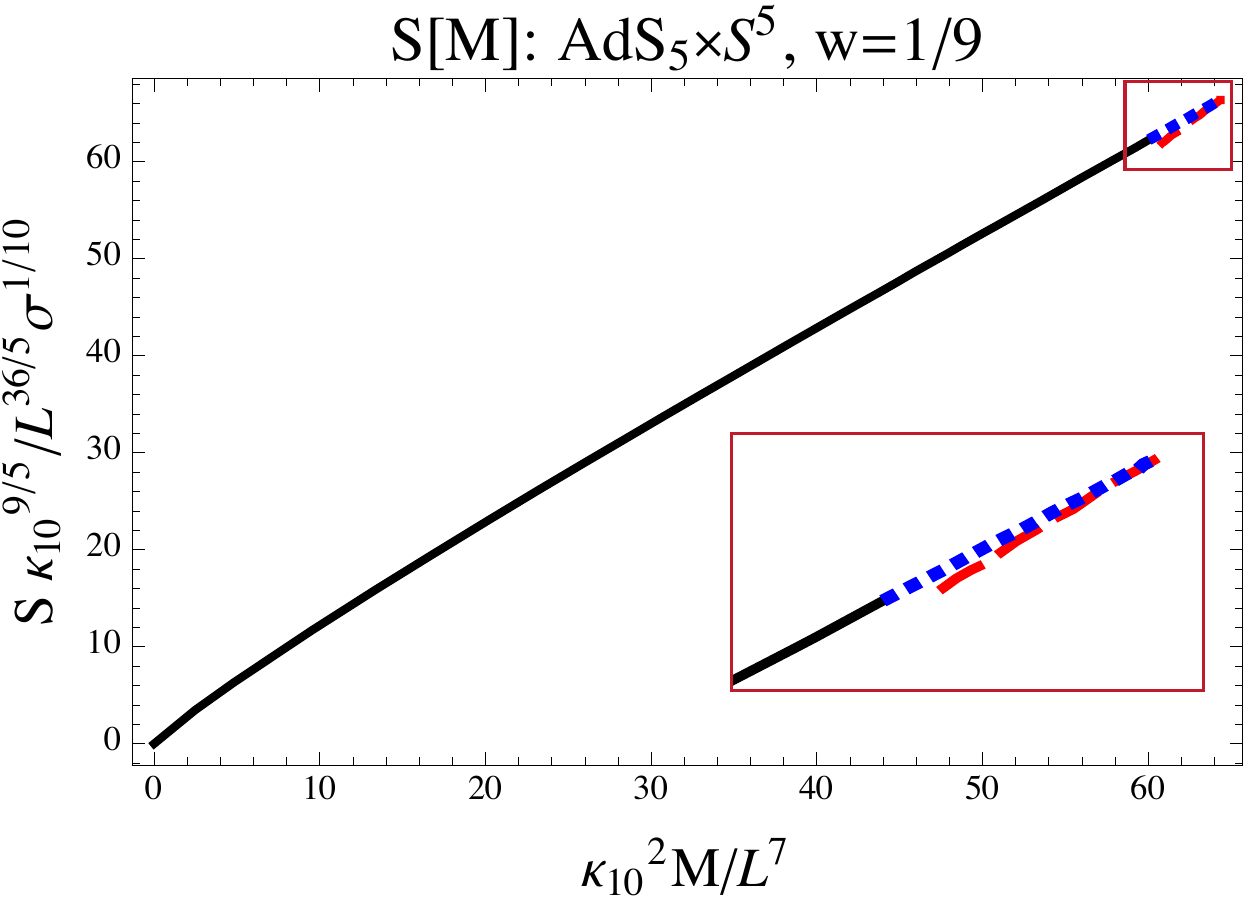}
\includegraphics[scale=.5]{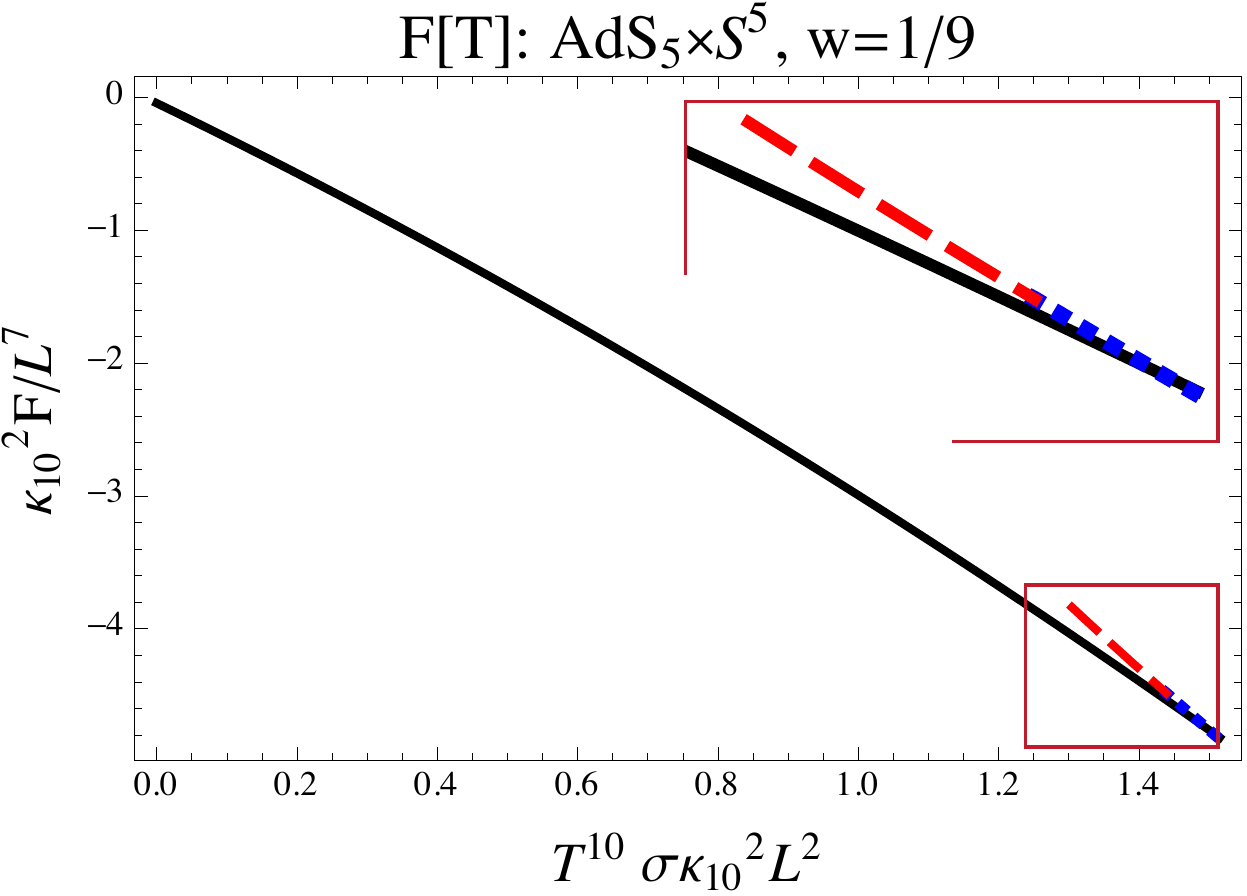}
\caption{
The energy-entropy curve and temperature-free energy curves for stars in $AdS_5\times S_5$. The solid black curve is both entropically and thermodynamically preferred, the blue dotted section is entropically preferred but thermodynamically disfavored, and the red dashed line is always disfavored. The structure is the same as for the $AdS_3$ and $AdS_5$ case, figures \ref{fig:ads3_mass_star} and \ref{fig:ads5_mass_star}. Note that $\kappa^2$ here is the {\it ten}-dimensional Planck scale.
\label{fig:ads5s5_mass_star}
}
\end{center}
\end{figure}

We will demand that the asymptotic behavior as $r \to \infty$ for solutions to (\ref{eq:FR_rho_eom}) - (\ref{eq:FR_C_eom}) is:
\be
f=\frac{r^2}{L^2}+1-\frac{\epsilon}{r^2}+\ldots,~\rho=\frac{\mu}{r^{10}}+\ldots~\chi=\frac{\chi_0}{r^8}+\ldots,~C=\frac{\chi_0}{30 r^8}+\ldots\ ,
\ee
while the behavior at the origin is:
\be
f=1+\Op(r^2),~\rho=\rho_{core}+\Op(r^2),~\chi=\chi_{core}+\Op(r^2),~C=C_{core}+\Op(r^2).
\ee
This demand is nontrivial for $C$; it is a massive field in $AdS_5$, and so there is a potentially non-normalizable solution which can grow rapidly at large radius, whose backreaction destroys the asymptotic AdS structure.  This leads to a  delicate shooting problem when we fix the boundary conditions $\rho(0),~C(0)$ at the origin: a small error leads to a large non-normalizable component for $C(r)$ at large radius. In the end, there remains a one-parameter family of asymptotically $AdS_5\times S_5$ solutions which are spherically symmetric in both factors. We find that the radius of the $S_5$ varies only weakly in $r$. If we measure the size change by $L_{S_5}(r=0)/L_{S_5}(r=\infty)$, we see that even at core densities beyond the unstable point, the radius of the $S_5$ only changes by a few percent (see figure \ref{fig:ads5s5_radius_change}).
 We show the $S(E)$ and $F(T)$ curves in Fig. (\ref{fig:ads5s5_mass_star}), and find that the qualitative results of  Section $\ref{subsec:AdS5}$ indeed still hold.  The one difference is the scaling of the temperature at which the canonical ensemble becomes unstable: it occurs when $\sigma_{10} (LT)^{10} \sim L^8/\kappa_{1}^2 \sim 1/N^2$.

Because our analysis is restricted to $s$-wave configurations on the sphere, an open question is whether an inhomogeneous mode on the sphere could develop an instability before the homogeneous mode does.

\subsubsection{$AdS_{3} \times S_{3}$ stars}

Starting with type II supergravity in 10 dimensions and reducing on $T^4$, we can consistently truncate  to $\mathcal{N}=1,~D=6$ supergravity, which has a metric, self-dual 3-form field strength, and complex Weyl spinor satisfying $\Gamma_7\psi_+=+\psi_+$ \cite{Marcus:1982yu,Nishino:1984gk}\footnote{As usual we can't write down a consistent Lorentz invariant Lagrangian for the self-dual form. This problem can be alleviated by adding a antisymmetric tensor multiplet with an anti-self dual field strength, a complex Weyl spinor satisfying $\Gamma_7 \psi_- = -\psi_-$ \cite{Marcus:1982yu}. However, when constructing perfect fluid stars we need only concern ourselves with the bosonic equations of motion.}. The equations of motion are 

\be
R_{ab} = H_{acd}H_{b}{}^{cd},~H_{abc} = \frac{1}{3!}\epsilon_{abcdef}H^{def}.
\ee
Just as in \eqref{eq:ads5s5metric}, we use the ansatz
\be
ds^2 =ds^2_3+L^2 e^{2C}d\Omega_3^2,~F=L^2(1+*)\epsilon_3.
\ee

Coupling to our perfect fluid \eqref{fluidtensor}, we have
\be
R_{ab} = H_{acd}H_b{}^{cd} + \kappa_6^2 \{T_{ab} - g_{ab} T^c{}_c/4\},
\ee
This and conservation gives the following equations
\be
\frac{(2+3 r C')p_{,\rho}}{p+\rho}\rho'-3 C' - \frac{3}{2} r C'^2 + \frac{2r}{f L^2}\left(3e^{-C}-2 e^{-3 C} + \kappa^2L^2p \right)=0,\label{eq:FR_s_rho}
\ee
\be
(2+3 r C')f'+6 r f C'^2 - \frac{4r}{L^2}(4e^{-3C}-3e^{-C})+\kappa_6^2 r (\rho+3p) +\frac{3r^2\kappa_6^2 C'}{2}\left(\rho-p-\frac{8e^{-3C}}{\kappa_6^2 L^2} \right) =0,\label{eq:FR_s_f}
\ee
\be
(2+3 r C')\chi' - 6 C' + 3 r C'^2 - \frac{12 r e^{-3 C}}{L^2 f}(2-2e^{2C}+r C')+\frac{r \kappa_6^2}{2f}\left[ p(14-3r C')+\rho(2+3r C')\right]=0,\label{eq:FR_s_chi}
\ee
\be
C'' - \frac{\kappa_6^2 (p-\rho)}{2f}-\frac{4}{L^2 f} (e^{-C}-e^{-3C})+\left(\frac{1}{r}+\frac{2re^{-3C}}{L^2f}+\frac{\kappa_6^2 r (p-\rho)}{4f} \right)C'=0.\label{eq:FR_s_c}
\ee
In what follows we will focus on the case of linear equations of state, $p(\rho) = w \rho$, and not just $w=1/5$ for reasons that will become clear shortly. At large radius, we have
\be
\rho=\frac{\mu}{r^\frac{1+w}{w}}+\ldots,~
f=\frac{r^2}{L^2}-\epsilon+\ldots,~
\chi=\chi_0/r^4+\ldots,~C=\chi_0/15r^4+\ldots.
\ee
\begin{figure}[thbp]
\begin{center}
\includegraphics[scale=.5]{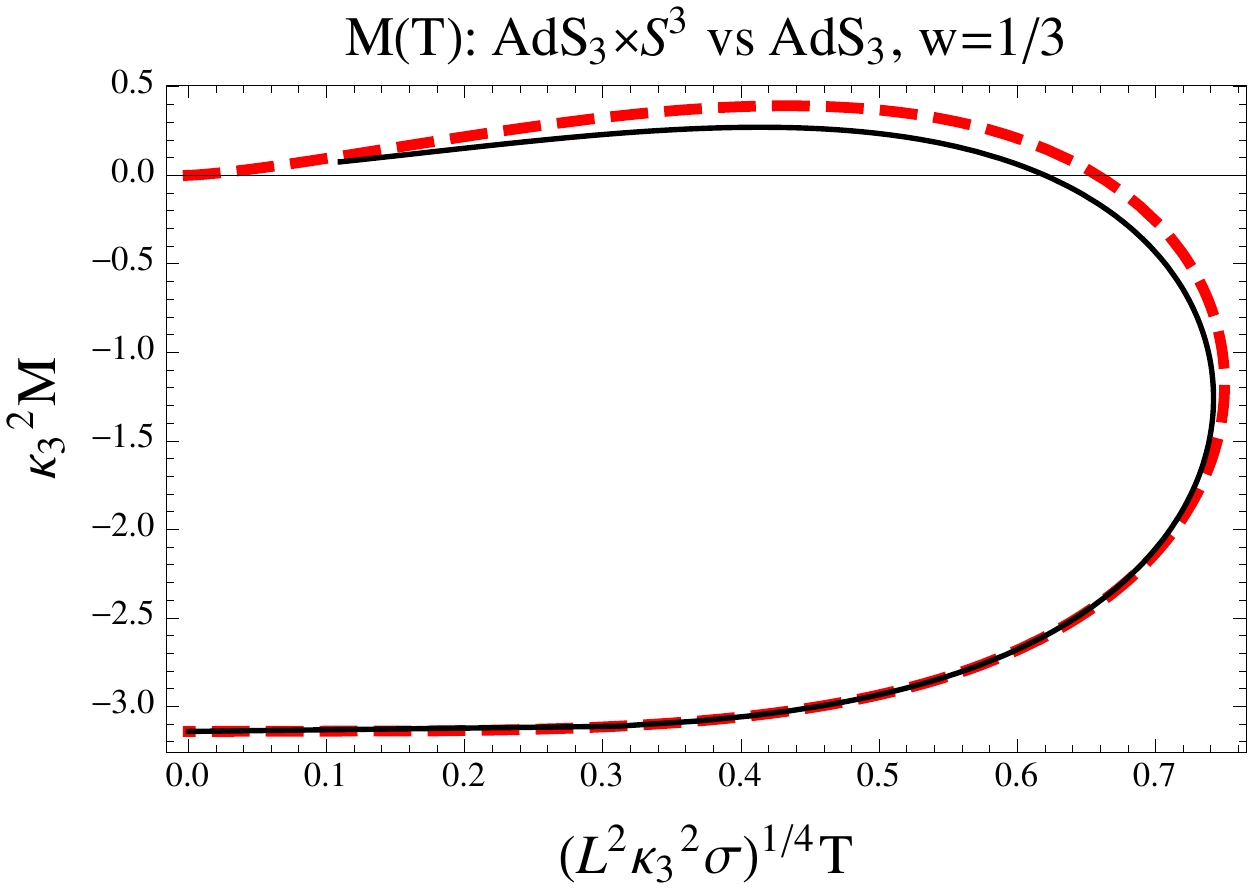}
\includegraphics[scale=.5]{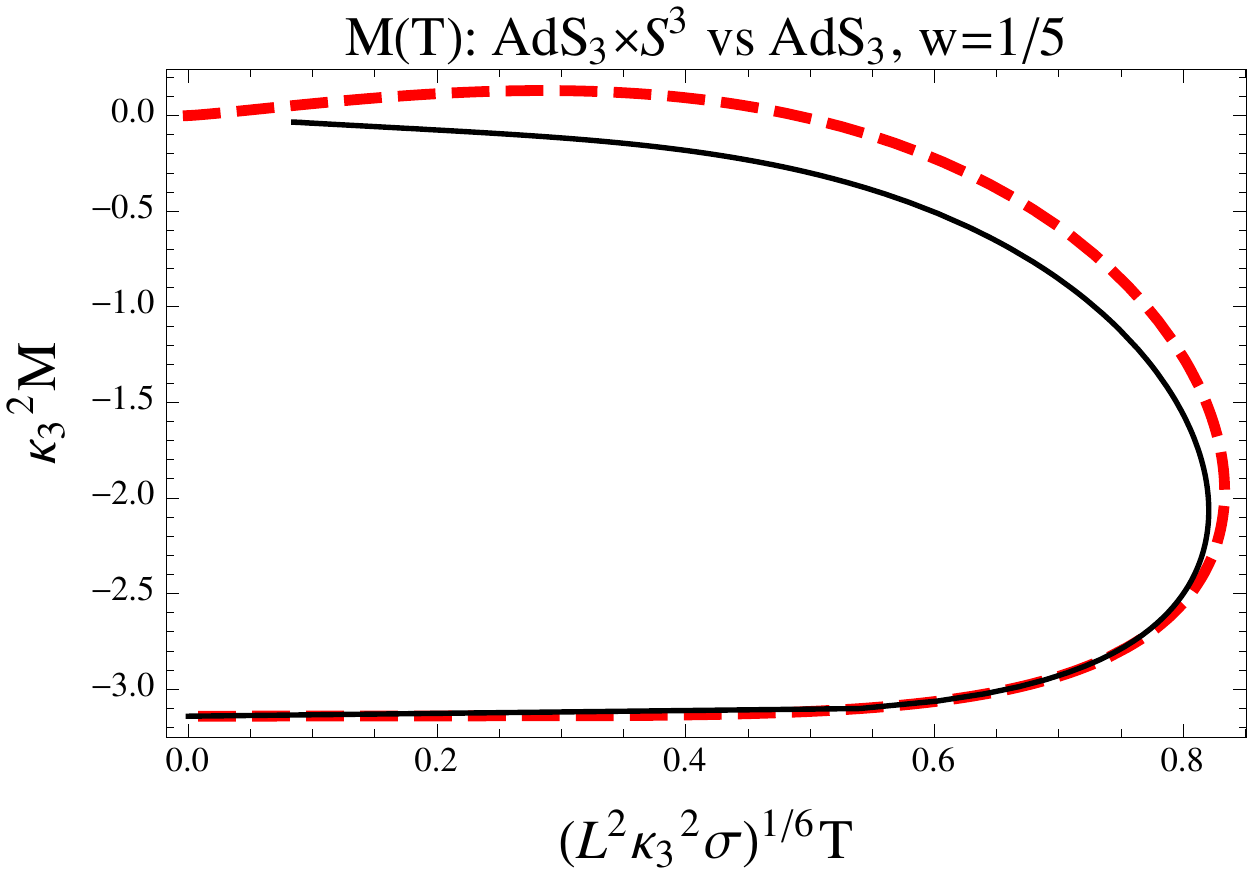}
\caption{Comparison of $AdS_3\times S^3$ numerically constructed stars to $AdS_3$ analytic solutions. Note that the dilaton only mildly changes the structure of the solutions, though the effect is clearly stronger for smaller $w$. For the case of 6D radiation we are unable to numerically find the mass turning point, though we conjecture it exists.
\label{fig:ads3_MT_plots}}
\end{center}
\end{figure}
Unfortunately, due to the dilaton we can not find explicit analytic solutions. Constructing solutions numerically (and treating the massive dilaton in the same way as we did for the $AdS_5\times S^5$ case) we find structure of the solutions is qualitatively extremely close to that of the analytic $AdS_3$ system. In figure \ref{fig:ads3_MT_plots} we plot energy versus temperature for $w=1/5$ and $w=1/3$ for $AdS_3\times S^3$ alongside the analytic solution (\ref{eq:ads3mass}, \ref{eq:ads3tempent}). When studying the case of 6D radiation, we are unable to find the mass turning point  as it appears to occur at a very high core density where we lose numerical control. However the shape is remarkably similar to the $AdS_3,~w=1/5$ solution. We conjecture that the mass turning point still exists and the phase diagram is the same as for $AdS_3$. For the case of $w=1/3$ the mass turning point occurs sooner (that is, at a lower core density) and again we find qualitative agreement with the $AdS_3$ system and therefore the results of section \ref{subsec:AdS3} still hold.

\section{Thermodynamics and the gauge theory dual}


In this section we map out the thermodynamic landscape of gravitational theories in asymptotically AdS spacetimes, paying attention to the implications for the gauge theory duals.

In the prior section we focused on the self-gravitating radiation, including the entropy and free energy of these solutions.  These configurations are known to dominate the thermodynamics at low temperatures and energies.  Gravity at high energes and temperatures in AdS is known to be dominated by black holes, with large entropies of order $R^{D-2}/\kappa_D^2$ for $AdS_D$ backgrounds \cite{Hawking:1982dh}. In the limit of large Planck mass, there is a sharp first-order phase transition at finite temperature. When there is a gauge theory dual, this represents a deconfinement transition, with a large number of degrees of freedom participating in the thermodynamics \cite{Witten:1998qj,Witten:1998zw}.

For systems in infinite volume with sufficiently short-range interactions, the canonical and microcanonical ensembles are known to be equivalent.  However, the above transition occurs when the gauge theory is at finite volume, and is dominated by spherically symmetric configurations. In such cases the ensembles are known to be generically inequivalent.  A sign of this is the presence of microcanonically stable black holes with negative specific heat \cite{Banks:1998dd,Horowitz:1999uv}.  Nonetheless, the microcanonical phases can be identified with equilibrium points in the canonical ensemble, although these may be metastable or unstable points.  

We will open by reviewing some basic aspects of first-order transitions in systems with inequivalent microcanonical and canonical ensembles.  Following this, we will explore in detail the landscape of $AdS_5\times S_5$ and $AdS_3\times S_3$ compactifications.  Finally, we will discuss the implications of this landscape for the dynamics of small perturbations at finite temperature, which has been discussed as a diagnostic of unitarity of black hole evolution \cite{Maldacena:2001kr}.

\subsection{Review: microcanonical versus canonical thermodynamics}

\begin{figure}
	\centering
	\begin{subfigure}[h]{0.45\textwidth}
		\centering
		\includegraphics[width=\textwidth]{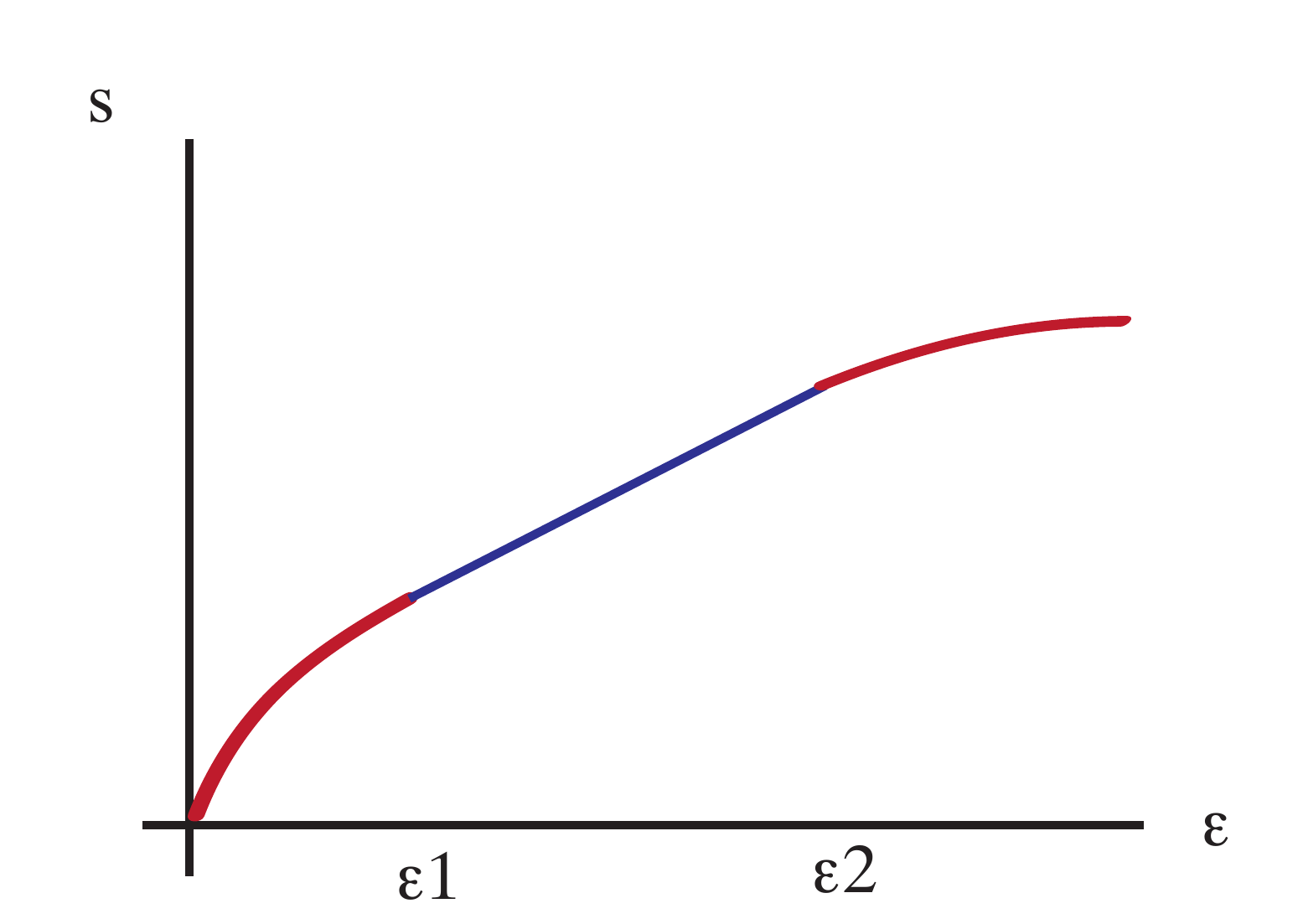}
		\caption{Microcanonical phase curve}
		\label{fig:MCTL}
	\end{subfigure}
	\begin{subfigure}[h]{0.45\textwidth}
		\centering
		\includegraphics[width=\textwidth]{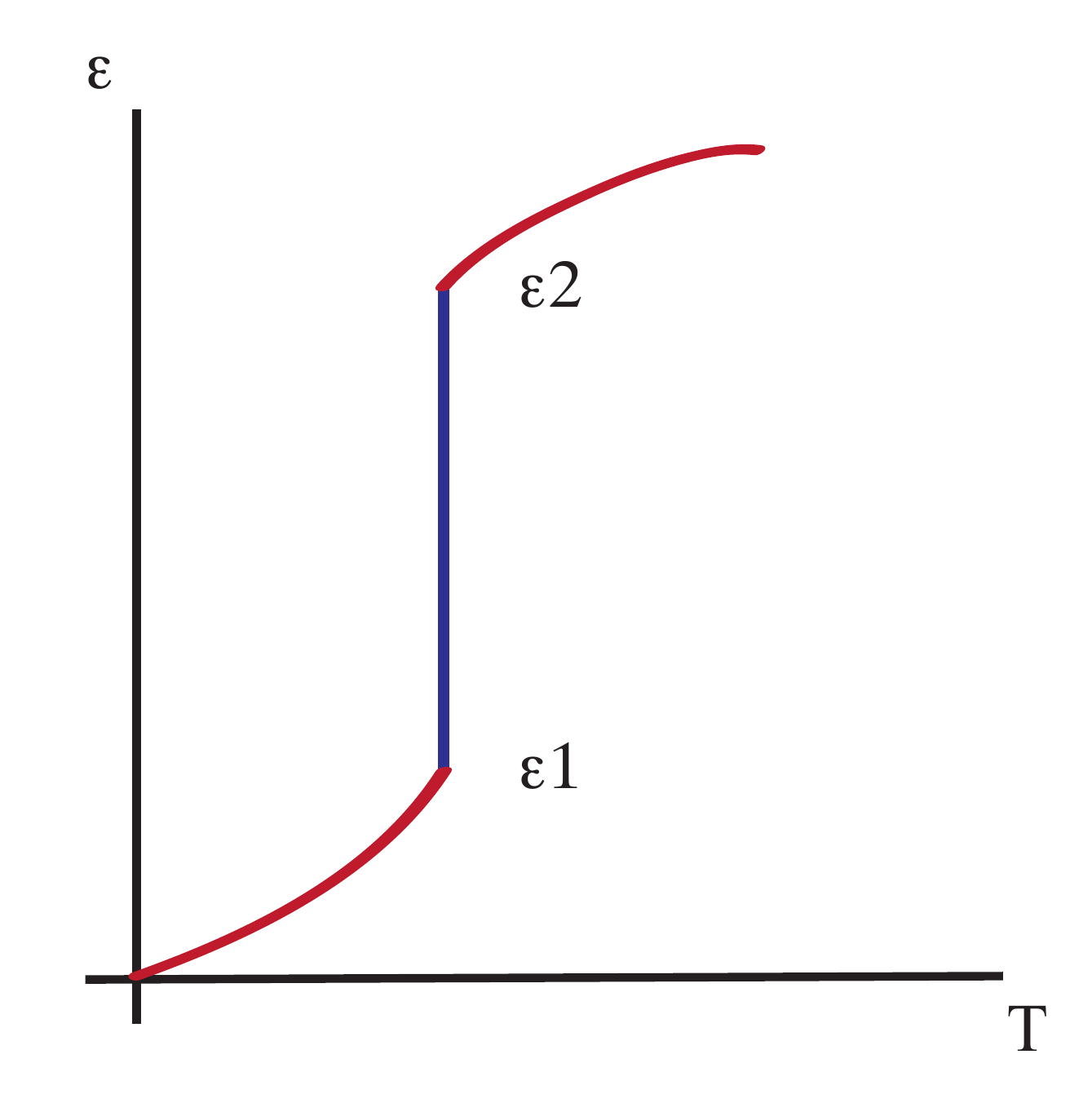}
		\caption{Canonical phase curve}
		\label{fig:CTL}
	\end{subfigure}
	\caption{Phase curves for a system with a first-order transition and short-range interactions in the thermodynamic limit. 		Figure (a):  Entropy density ($s$) as a function of energy density ($\epsilon$) in the microcanonical ensemble. Figure (b): 		Energy density as a function of temperature in the canonical ensemble}
	\label{fig:FirstOrderMC}
\end{figure}

For local field theories with short-range forces, the microcanonical and canonical ensembles are equivalent in the thermodynamic limit. This ensures that the specific heat at constant volume is positive at all temperatures and energies.  For theories of this kind with a first order phase transition, the curves $s(\epsilon)$ and $\epsilon(T)$ are shown in Figure \ref{fig:FirstOrderMC}. In this figure $s,\epsilon$ are the energy and entropy per unit volume, and $T = \left(\frac{\p s}{\p \eps}\right)^{-1}$.  In Figure \ref{fig:MCTL}, the curve corresponds to the configurations with maximum entropy at fixed energy; in Figure \ref{fig:CTL}, the curves correspond to configurations with maximum free energy at fixed temperature.  The straight line in the curve $s(\epsilon)$ is a coexistence line, in which a fraction $f_{high} = \frac{\eps - \eps_1}{\eps_2 - \eps_1}$ of the total system consists of domains in the high energy phase with energy density $\eps_2$. The domain boundaries contribute due to the surface tension, but their contribution is subextensive in the thermodynamic limit. All the points on this line have the same temperature.  The result is the sudden discontinuous jump in energy in the curve $T(\eps)$, which signals a first order phase transition.  Note also that the specific heat $C_V = \p E/\p T$ has a delta function discontinuity here.

\begin{figure}
	\centering
	\begin{subfigure}[h]{0.45\textwidth}
		\centering
		\includegraphics[width=\textwidth]{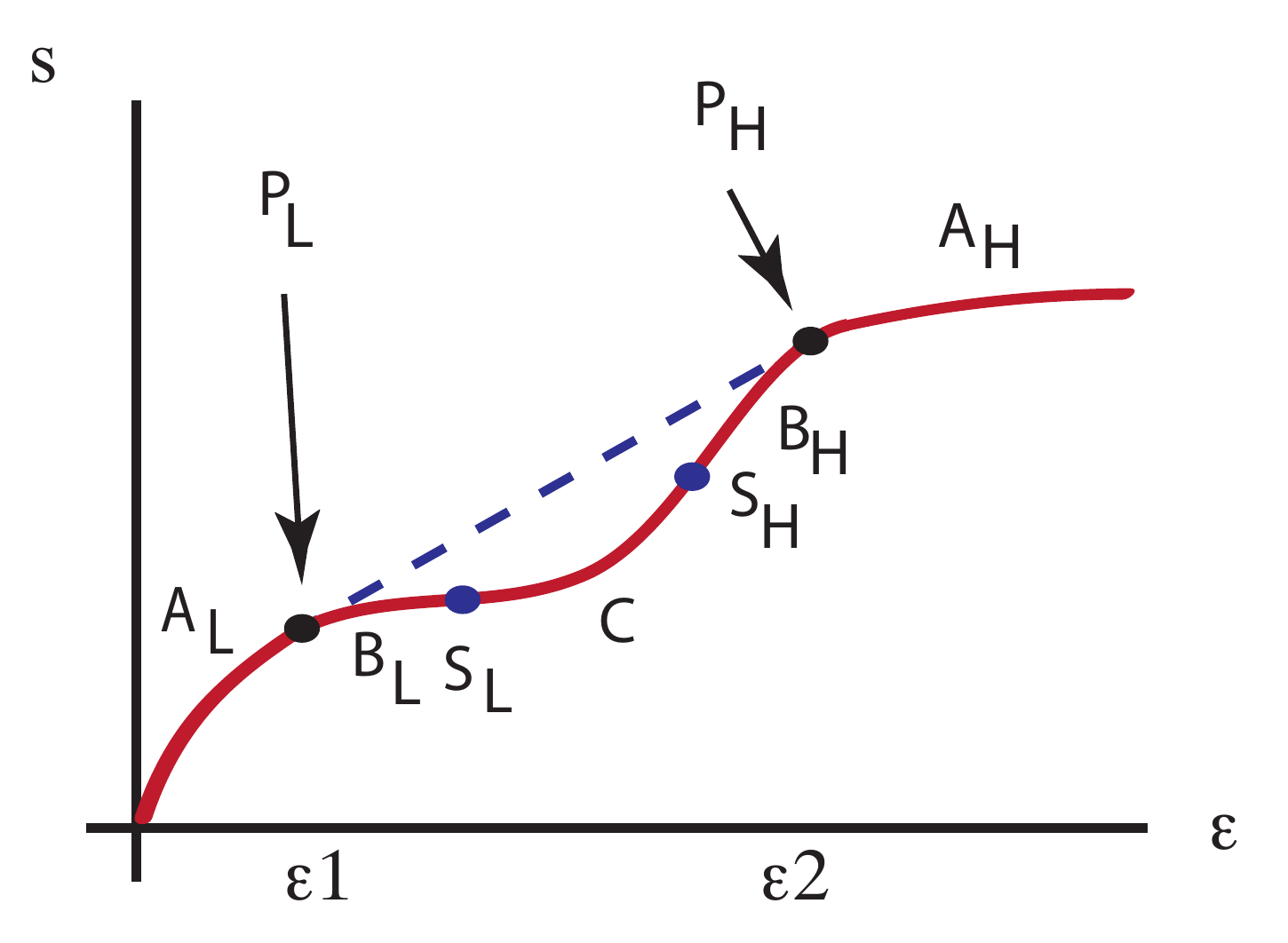}
		\caption{Microcanonical phase curve}
		\label{fig:MCNE}
	\end{subfigure}
	\begin{subfigure}[h]{0.45\textwidth}
		\centering
		\includegraphics[width=\textwidth]{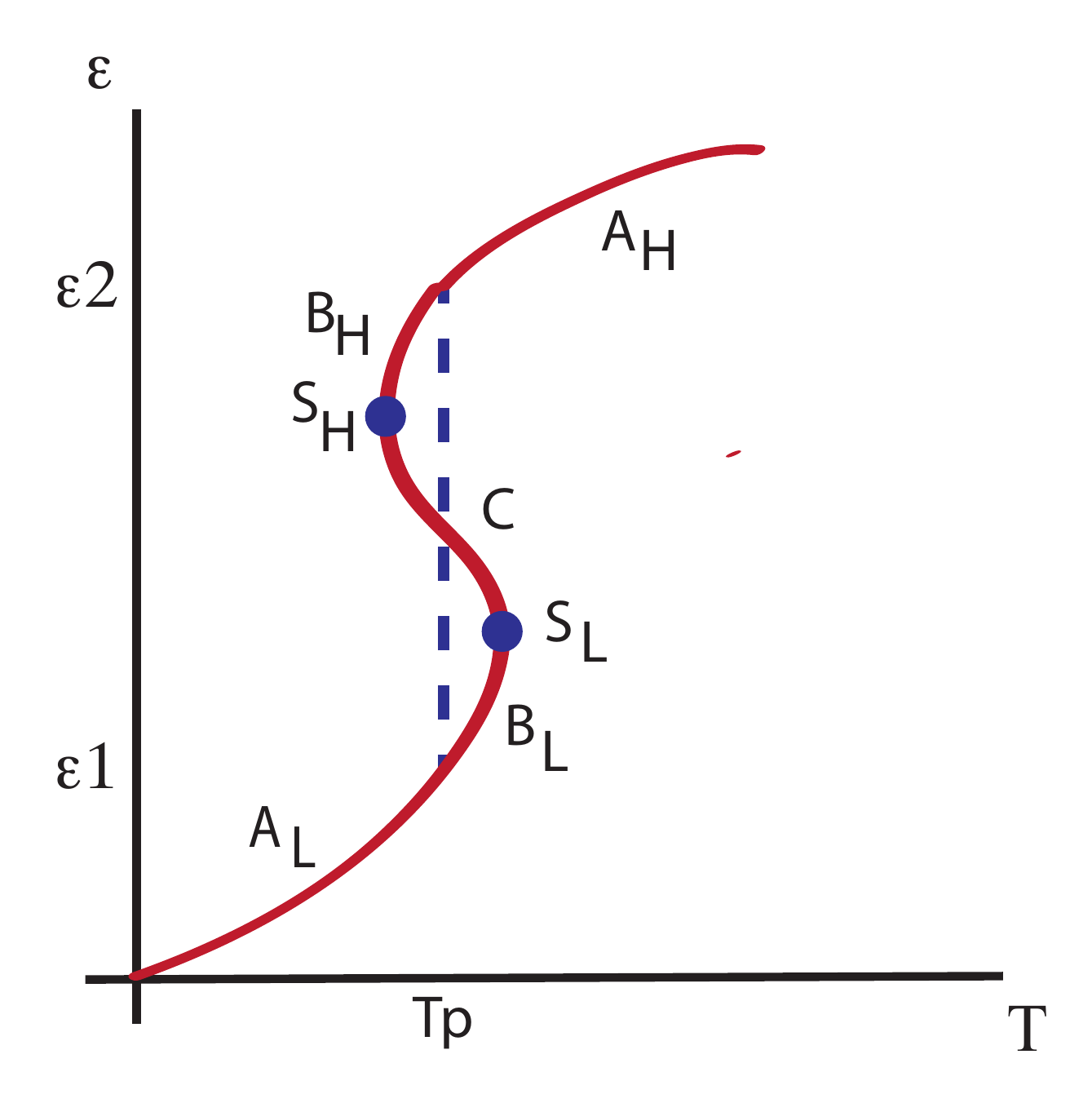}
		\caption{Canonical phase curve}
		\label{fig:CNE}
	\end{subfigure}
	\caption{Phase curves for a a system with a first-order transition, when the canonical and microcanonical ensembles 	are not equivalent. Figure (a) denotes the curve in the microcanonical ensemble, and figure (b) is the energy as a 	
	function of microcanonical temperature. The regime $\eps_1 < \eps < \eps_2$ is the coexistence region.  
	The regions $B_{L,H}$ are local but not global minima of the free energy, and have positive specific heat.  The regime $C$ (the ``convex intruder'') has negative specific heat; when there is a set of macrostate variables for each energy, it will be a saddle point or local maximum of the free energy. Finally, the points $P_{L,H}$ correspond to the two phases at the phase transition temperature $T_p = (\p S/\p E)^{-1}\big|_{P_{L,H}}$.}
	\label{fig:FirstOrderNE}
\end{figure}

However, there are many cases in which the two ensembles are known to be inequivalent.  This generically happens for finite systems away from the thermodynamic limit; it can also occur for theories with sufficiently long-range forces such as mean field theories or gravitating ensembles, in which the interaction between phase domains effectively no longer occurs just along the domain boundaries.  In these cases, the $s(\epsilon)$ curve for a theory with a first-order phase transition  as shown in Figure \ref{fig:FirstOrderNE} \cite{huller1994first}.  The coexistence line bends in and includes a  piece known as a ``convex intruder''.  A heuristic explanation of this for local theories is that at finite volume, the surface tension contributes a finite amount to the energy of a given configuration of phase domains \cite{gross1996microcanonical,2000EPJB...15..115G}.   This structure is characteristic of a probability distribution with two distinct peaks at different energies \cite{wales1994coexistence}. In this case, there is a region where the entropically dominant configuration in the microcanonical ensemble has negative specific heat, and corresponds to a thermodynamically unstable configuration in the canonical ensemble.

When there is a set of macroscopic variables $\phi$ characterizing the phase at each value of the energy, and there is a good Landau free energy that can be written as a function of $\phi$, each part of the $s(\eps)$ curve has an interpretation in the canonical ensemble even when the ensembles are inequivalent \cite{touchette2005nonequivalent}.  In Figure \ref{fig:FirstOrderNE} we show both the $s(\eps)$ and corresponding $\eps(T)$ curves in this case, where $T$ is defined as the microcanonical temperature. The region $A_L$ corresponds to the ``low-temperature'' phase when it is a global minimum of the free energy.  The two points $P_{L,H}$, which have the same slope (and thus the same temperature $T_p$) correspond to the high and low-temperature phases at the phase transition. The region $B_L$ corresponds to the ``low temperature'' phase at a temperature where it is metastable in the canonical ensemble.  The region $A_H$ corresponds to the high temperature phase when it is a global minimum of the free energy. The region $B_H$ corresponds to the ``high-temperature'' phase when it is a local minimum of the free energy, metastable in the canonical ensemble. The region C, with negative specific heat, corresponds to the local maximum of the free energy between the two phases.  The points $S_{L,H}$ correspond in the canonical ensemble to the ``spinodal'' points at which the local, metastable minima of the free energy coalesce with the local maximum and become unstable.

\begin{figure}
\vspace*{-1cm}
	\centering
	\begin{subfigure}[h]{0.45\textwidth}
		\centering
		\includegraphics[width=\textwidth]{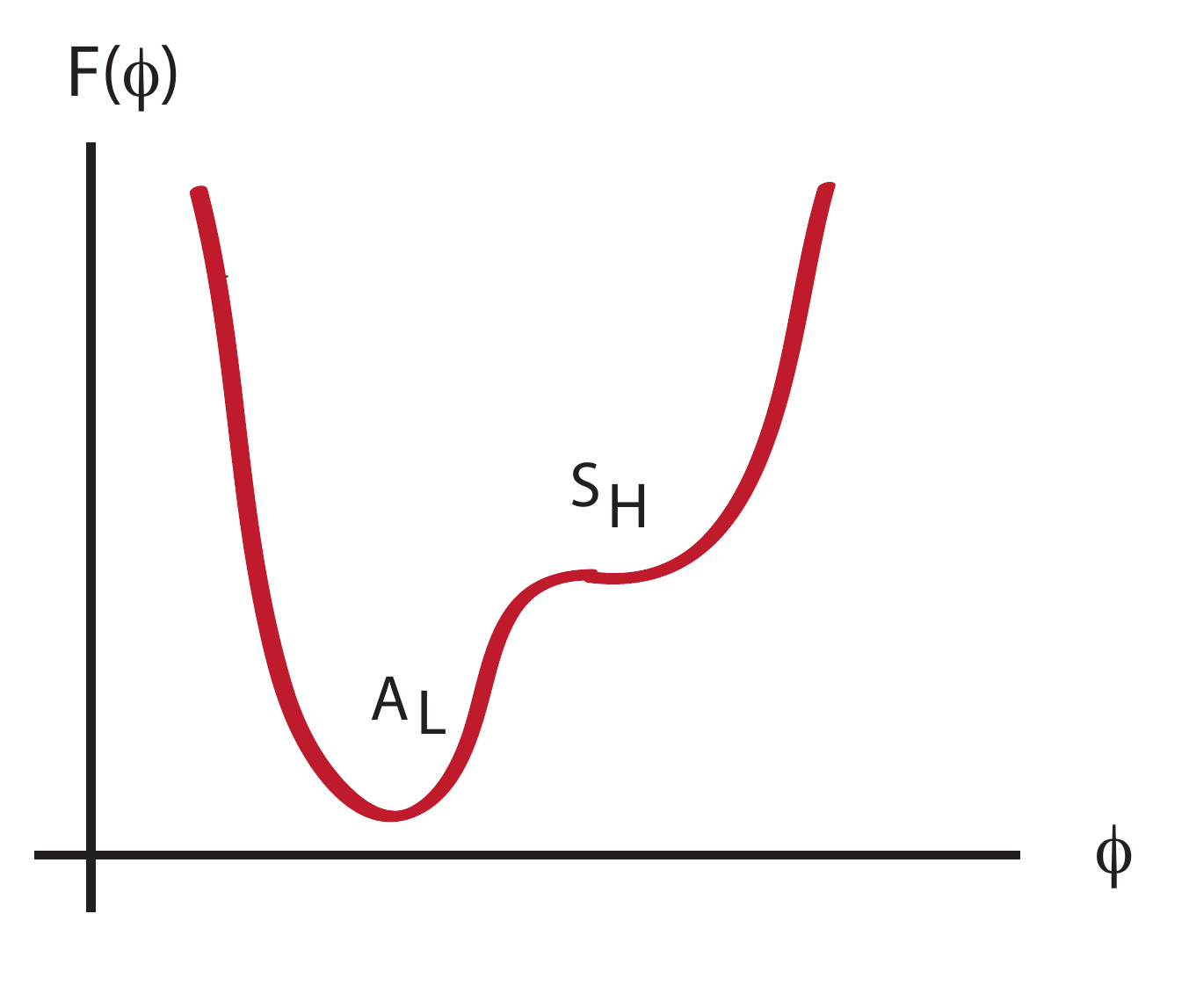}
		\caption{$T = T_{spinodal,H} = \left(\frac{\p s}{\p \eps}\right)\big|_{\eps_{S_H}}$}
		\label{fig:CSPL}
	\end{subfigure}
	\begin{subfigure}[h]{.45\textwidth}
		\centering
		\includegraphics[width=\textwidth]{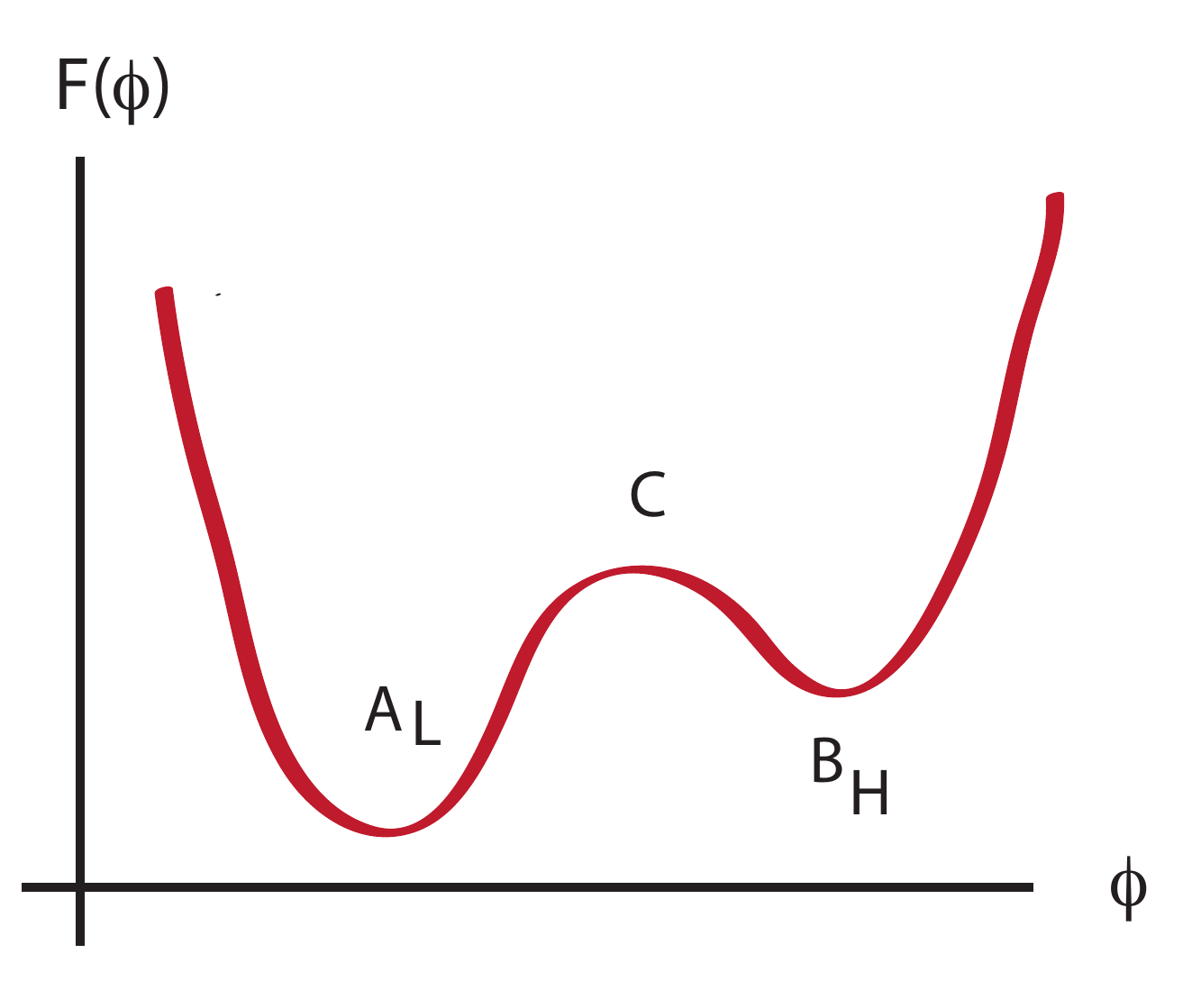}
		\caption{$T_{spinodal,H} < T < T_p$}
		\label{fig:CLT}
	\end{subfigure}\\
	\begin{subfigure}[h]{.45\textwidth}
		\centering
		\includegraphics[width=\textwidth]{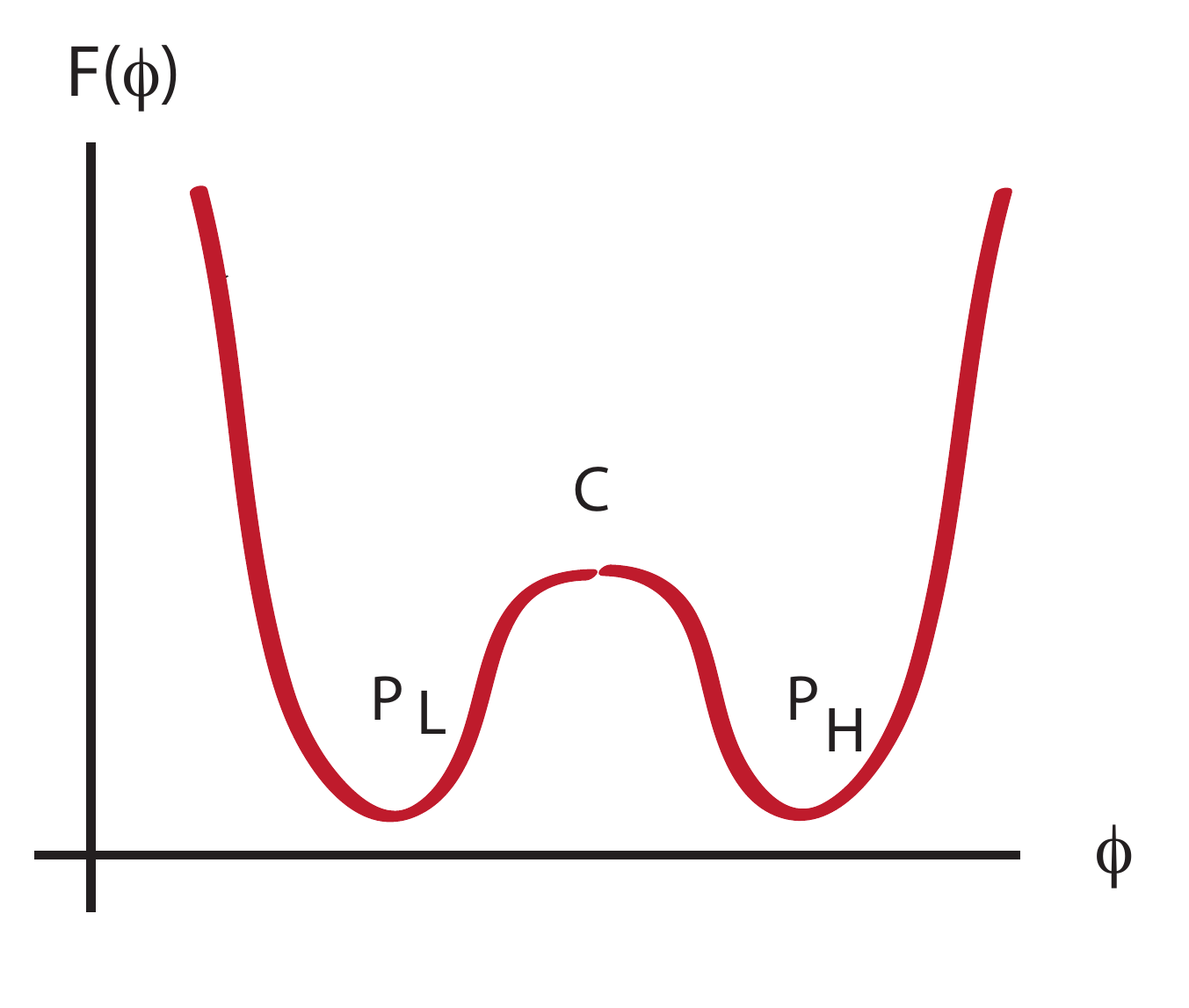}
		\caption{$T = T_p$}
		\label{fig:CPT}
	\end{subfigure}
	\begin{subfigure}[h]{0.45\textwidth}
		\centering
		\includegraphics[width=\textwidth]{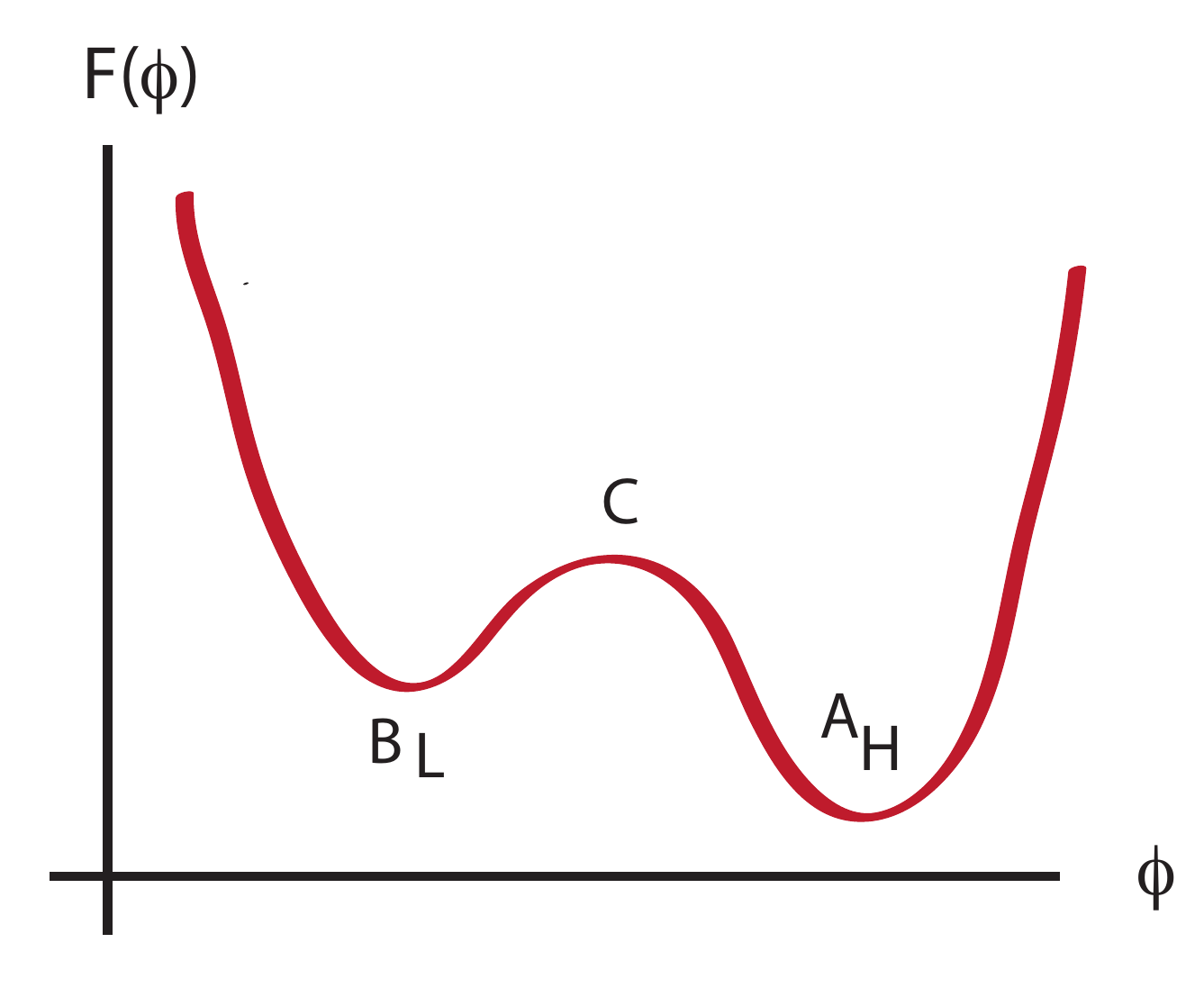}
		\caption{$T_p < T < T_{spinodal, L} = \left(\frac{\p s}{\p \eps}\right)\big|_{\eps_{S_L}}$}
		\label{fig:CHT}
	\end{subfigure}\\
	\begin{subfigure}[h]{0.45\textwidth}
		\centering
		\includegraphics[width=\textwidth]{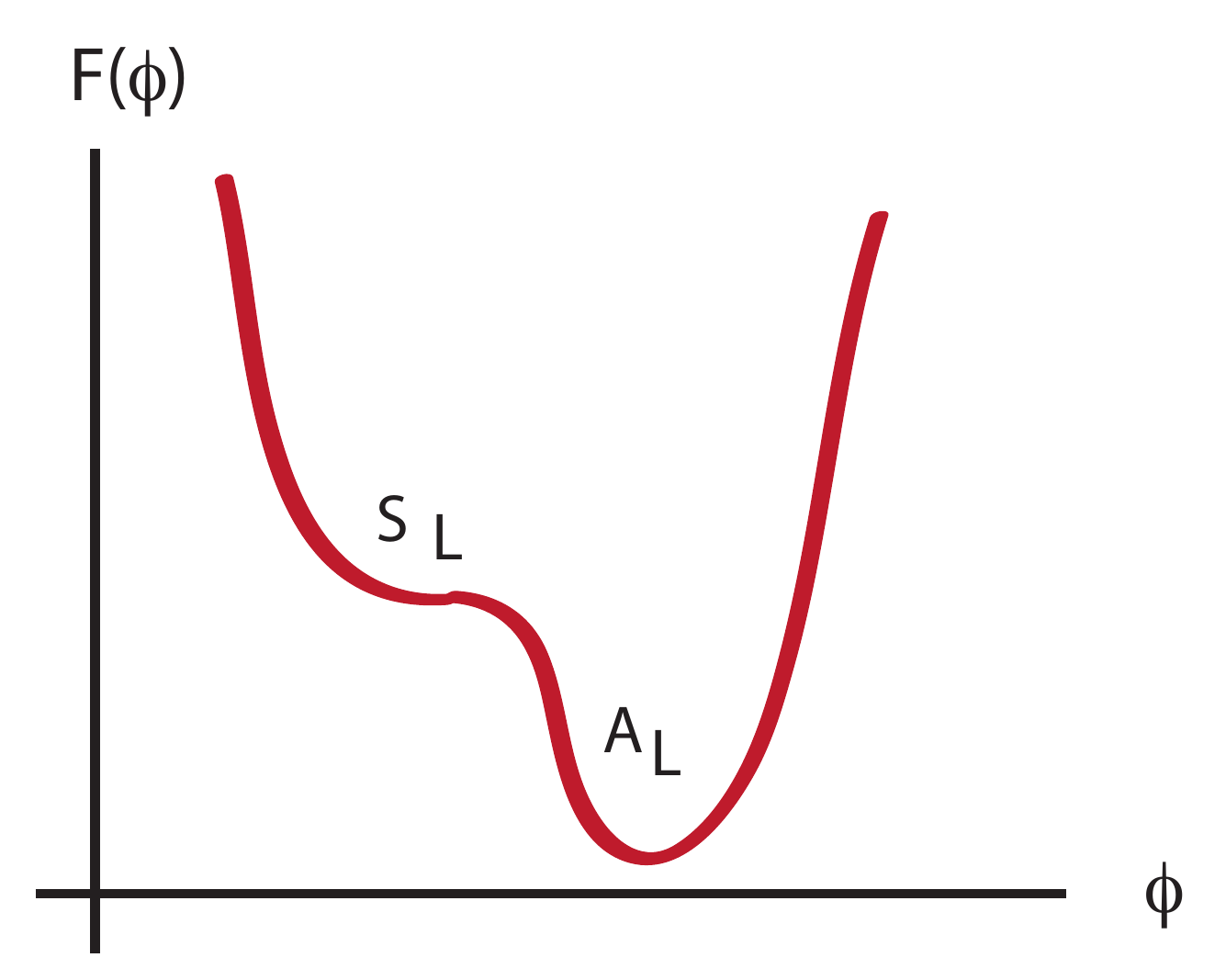}
		\caption{$T = T_{spinodal, L}$}
	\end{subfigure}
	\caption{  Canonical free energies as a function of an order parameter $\phi$, as one dials the temperature through the 
	first-order phase transition. Extrema of the free energy are labeled by the corresponding phases on the microcanonical curve,
	in the case that the ensembles are not equivalent.}
	  \label{fig:FirstOrderCEPT}
\end{figure}

For the case of a single variable $\phi$, Figure \ref{fig:FirstOrderCEPT} shows the canonical free energy curves at various temperatures as a function of $\phi$, and identifies the extrema with the microcanonical phases. This shows a simple scenario.  More generally there can be additional saddle points or maxima of the canonical free energy, as a function of the parameters $\phi$, which do not correspond to any microcanonical phase.\footnote{We would like to thank Hugo Touchette for explaining this to us} In particular, we will argue below that this occurs for $2d$ CFTs with gravitational duals.

\subsection{$AdS_5\times S_5$ and $d=4$, ${\cal N} = 4$ large-N super-Yang Mills}

We begin by considering maximally supersymmetric type IIB string theory on $AdS_5\times S_5$, with the $AdS_5$ factor written in global coordinates; this is dual to ${\cal N}= 4$ super-Yang-Mills theory on $S_3\times R$, and it has a first-order phase transition at finite temperature \cite{Hawking:1982dh} above which the theory deconfines \cite{Witten:1998qj,Witten:1998zw}. The specific values of the temperature and energy at the phase transitions are of course known, and are reviewed in Appendix \ref{app:conventions}. 

\subsubsection{Review of phase structure}

In the standard scaling limit used in the AdS/CFT correspondence, in which we take $N \to \infty$ with $\lambda$ fixed and large, the microcanonical and canonical phase structures have been discussed in a number of publications \cite{Banks:1998dd,Horowitz:1999uv,Barbon:2001di,Aharony:2003sx,Aharony:2005bq}.

The different microcanonical phases are:
\begin{enumerate}
\item A phase consisting of a hot gas of supergravity modes.  This will have entropy $S = \sigma (E L)^{4/5}$ for some constant $\sigma$ proportional to the Stefan-Boltzmann constant, for $E L < 1$.  For higher energies, $S = \sigma_{10} (E L)^{9/10}$, characteristic of a ten-dimensional gas. This  corresponds to phases $A_L$, $P_L$, and $B_L$ in Figure \ref{fig:FirstOrderNE}.
\item When $(E L) \geq \lambda^{5/2}$ there is a ``Hagedorn'' phase dominated by a long string \cite{Mitchell:1987hr,Mitchell:1987th,Bowick:1989us}, with entropy $S = 4\pi \sqrt{\alpha'} E - a \ln E$.  Here $a \geq 1$ will have a soft dependence on $E$, depending on the size of the string with respect to the AdS scale. Note that these states have negative specific heat due to the logarithmic term; nonetheless they can be stable in the microcanonical ensemble.  This phase  corresponds to $S_L$ and part of $C$ in Figure \ref{fig:FirstOrderNE}. Note that in this energy range, the self-gravitating gas is still a mechanically stable configuration, but with  subdominant entropy
\item $10d$ ``small black holes'', Schwarzschild black holes with horizon size less than $L$, localized on the $S_5$.  These have negative specific heat, but can be stable in the microcanonical ensemble, when they equilibrate with their Hawking radiation. \cite{Hawking:1982dh,Horowitz:1999uv}. These black holes maximize the entropy when $E L \geq N^{20/17}$ \cite{Horowitz:1999uv}, and correspond to much of region C in Figure \ref{fig:FirstOrderNE}. We estimate that the top end of this range is at $(EL) \sim .1 N$, at which point the 5d black holes begin to dominate the thermodynamics.  Based on the Gregory-LaFlamme-like instabilities of the 5d black hole \cite{Hubeny:2002xn}, we expect this phase to extend at least up to energies of order $EL \sim .2 N^2$. It may extend farther, but absent a specific solution it is hard to tell.  We also expect that the self-gravitating supergravity gas continues to be stable well into this regime, athough it is entropically subdominant.  
\item $5d$ ``small black holes'' (black hole solutions that are uniform on $S_5$) with horizon radius less than $L/\sqrt{2}$ and with negative specific heat, corresponding to the highest energy portion of phase C in Figure \ref{fig:FirstOrderNE}.  These are mechanically unstable to collapse to 10d black holes when $EL \sim .2 N^2$ \cite{Hubeny:2002xn}; their specific heat becomes positive when $EL = 9 N^2/16$, at the point $S_H$. 
Meanwhile, in the same energy regime, a second branch of supergravity solutions opens up at $EL \sim 20 L^3/{\kappa_5^2} \sim .5N^2$, as can be seen from \figref{fig:ads5_mass_star} of the previous section.  The branches meet at energies close to the end of the small black hole phase.
\item ``Large black holes'', 5d AdS black holes with positive specific heat.  These cover regions $B_H, P_H$, and $A_H$.  Note that the Hawking-Page temperature $T_{HP} = 3/(2\pi L)$ corresponds to the inverse slope of $s(\epsilon)$ at $P_H$, when $EL = 3 N^2/2$.
\end{enumerate}

We next revisit the canonical phase diagram.  The order parameter direction in the canonical phase curves in \figref{fig:FirstOrderCEPT}\ can be identified with a Polyakov loop-like variable \cite{Aharony:2005bq}. The canonical phases are:
\begin{enumerate}
\item Below the ``spinodal'' temperature $T_{S_H} = \left(\frac{\p S}{\p E}\right)^{-1}\big|_{S_L} = (\sqrt{2} \pi)/L$, the self-gravitating supergravity gas, corresponding to region $A_L$ of the microcanonical phase curve, is the global minimum and there are now other known local minima of the free energy. At $T = T_{S_L}$, an inflection point or ``spinodal'' appears corresponding to a black hole with (classically) infinite specific heat.
\item For temperatures $T_{S_L} < T < T_{HP} = 3/(2\pi L)$, the self-gravitating supergravity gas still describes the global minimum of the free energy; meanwhile, there is a metastable minimum of the free energy corresponding to a ``large'' AdS black hole with negative specific heat, and a maximum of the free energy correponding to a ``small'' AdS black hole with negative specific heat.  This is consisent with the results of \cite{touchette2005nonequivalent}, that the regime $B_H$ and $C$ in the microcanonical phase curve should correspond to a metastable local minimum and local maximum, respectively, of the free energy. 
\item At $T = T_{HP}$ the free energies of the supergravity gas and the ``large'' black hole agree, and the system undergoes a first-order phase transition. This corresponds to $P_L,H$ on the microcanonical phase curve.
\item For temperatures $T_{HP} < T < T_{S_L} \sim \lambda^{1/4}/L$, the large AdS black hole -- region $A_H$ on the microcanonical phase curve -- is a global minimum of the free energy; the self-gravitating supergravity gas is a metastable local minimum of the free energy, corresponding to region $B_L$ of the microcanonical phase curve, and the ``small'' AdS black hole is a local maximum or saddle point of the free energy, corresponding to region $C$ of the microcanonical phase curve.  Note that at temperatures slightly above the Hawking-Page temperature (about 4 percent), the 10d ``small'' black hole will begin to dominate. Again, these results are consistent with \cite{touchette2005nonequivalent}.
\item  At the temperature $T_{S_H}L \sim \lambda^{1/4}$, we reach another spinodal point at which the hot supergravity gas and the long string phase merge, corresponding to the Hagedorn transition in AdS space \cite{Barbon:2001di,Kruczenski:2005pj}; this is the ``correspondence point'' discussed in \cite{Horowitz:1996nw}. At higher temperatures, there is no metastable supergravity phase.
\end{enumerate} 

It would be interesting to find an interpretation of the small black hole phase as a ``coexistence phase'' in the dual gauge theory.  

\subsubsection{A different large-N limit and the appearance of the Jeans instability}

In the previous section, we found that no spherically symmetric configuration of hot self-gravitating matter exists above a critical energy of order $EL \sim L^3/G_N \sim N^2$.  At such energies, we expect the ``Jeans temperature'' to be $\sigma (T_{J}L)^5 \sim EL$.  To find the correct dependence, note that if $TL \gg 1$ we must think of the configuration as hot ten dimensional matter.  In this case,  we expect $\sigma_{10} (TL)^{10} \sim EL$, or $T_JL \sim N^{1/5}$.

If we dial $N \to \infty$ at fixed $\lambda$, $T_J L \gg T_{S,R} L \sim \lambda^{1/4}$, and the hot supergravity gas ceases to be a metastable configuration in the canonical ensemble at temperatures well below that at which the supergravity gas becomes relevant. This leads to the canonical phase structure described above. However, if we scale $N, \lambda$ so that $N^{1/5} \ll \lambda^{1/4}$, the supergravity gas ceases to exist as a configuration at energies well below this Hagedorn temperature.  This requires that $g_s \gg N^{-1/5}$.  Thus it is  possible to keep the string coupling weak and the Planck scale small. In this limit classical supergravity is still the leading approximation at large N, and our analysis is self-consistent.\footnote{However, if one attempts to compute string and quantum gravity corrections, these can mix; for example an ${\cal O}(\lambda^{-5/2})$ correction to classical supergravity will come in at the same order as the one-loop quantum gravity correction.}

In this scaling limit, the microcanonical phases simplify somewhat, as the Hagedorn phase is completely absent.  Instead, we expect region $B_L$ to extend to energies of order $N^{20/17}$, where the small 10d black hole (in equilibrium with its Hawking radiation) maximizes the entropy. The spinodal temperature, at which the small black hole and self-gravitating supergravity gas coalesce, is of order $T_{S_L}L \sim N^{2/17} \ll N^{1/5}$.  In the canonical ensemble, at temperatures slightly below $T_J$, we expect a new local saddle point to appear in the canonical free energy, corresponding to the second, thermodynamically unstable solution with higher core density.  We do not have a good description of an order parameter-like variable that interpolates to this solution.  At $T = T_J$, this saddle point coalesces with the metastable supergravity gas; above this temperature, only the large black hole is even metastable.

\subsection{$AdS_3\times S_3$ and $2d$ CFTs}

Next, we consider string theory on $AdS_3\times S_3\times M_4$, where $M_4$ is some Ricci-flat compact manifold such as $T^4$ or K3.  The solution can arise as the near-horizon limit of a bound state of NS5-branes and fundamental strings, a bound state of D5 and D1 branes, and so on. For specificity we will consider $M_4 = T^4$.  We will assume that the volume of $T^4$ is much smaller than the AdS radius, and that we are working at temperatures well below the inverse radius of the $T^4$, so that the physics is essentially six-dimensional.

If we have $p$ fundamental strings and $k$ NS5-branes, the bulk string and Planck scales are \cite{Giveon:1998ns,Maldacena:1998bw}:
\begin{eqnarray}
	\frac{8 \pi L}{\kappa_3^2} & = & 4 p k = \frac{2}{3} c \\
	\frac{L}{\ell_s} & = & \sqrt{k} 
\end{eqnarray}
where $\kappa_3^2$ is the 3-dimensional Planck scale, $\ell_s$ the string scale, $L$ the AdS radius of curvature, and $c = 6 kp$ is the central charge of the dual CFT. Similarly, for $p$ D-strings and $k$ D5-branes wrapping the $T^4$, we have:
\begin{eqnarray}
	\frac{8\pi L}{\kappa_3^2} & = & 4 p k = \frac{2}{3} c\\
	\frac{L}{\ell_s} & = & \left(\frac{kp}{v}\right)^{1/4} = \left(\frac{c}{6 v}\right)^{1/4}
\end{eqnarray}
Here $v = 1/g_6^2$, where $g_6$ is the string coupling in six dimensions; in the CFT dual to the D1-D5 system, it is an as yet unidentified modulus of the CFT.

\subsubsection{Review of phase structure}

The dual $2d$ CFT on a spatial circle of size $R$ has a first order Hawking-Page transition at finite temperature $TR = 1/2\pi$.  In the microcanonical ensemble, the phase structure is similar to that for $AdS_5$.  One difference is that there is no 3-dimensional black hole with negative specific heat. Once again, we work in the limit that $c \to\infty$ with $c/v$ fixed.

We will start by listing the microcanonical phases, using the D1-D5 description to discuss the gauge theory dual. Here we will set $\delta E = E + \frac{\pi}{\kappa_3^2}$, so that $\delta E = 0$ for the $AdS_3$ vacuum.
\begin{enumerate}
\item A phase containing (6d) supergravity modes, with entropy $S = \sigma (\delta E L)^{5/6}$. This will describe regions $A$, $P_L$, and part of $B_L$ in the microcanonical phase diagram.
\item When $\delta E \sim 1/\ell_s$, $EL \sim \left(\frac{c}{6 v}\right)^{1/4}$, a ``Hagedorn'' phase dominated by a single long string, again with $S = 4\pi \sqrt{\alpha'} \delta E - a \ln \delta E$, corresponding to $S_L$ and part of region $C$ in the microcanonical phase diagram. In this regime, the stable branch in \figref{fig:ads3_mass_star} continues to exist, but has lower entropy than the long string.
\item For $(1/\ell_s) < \delta E \lesssim \frac{\pi}{\kappa_3^3} + \Delta$, a phase corresponding to a 6d Schwarzschild black hole with horizon radius less than $L$. (The factor of $\Delta$ will be described in the next paragraph).  The entropy is
\be
	S \sim \left(\frac{9}{2c}\right)^{2/3} (E L)^{4/3}
\ee
and the specific heat is negative. This corresponds to part of $C$ in the microcanonical phase diagram, up to point $S_H$; there is no 3d black hole solution with negative specific heat.  Note that we do not have good solutions for this phase, especially close to $S_H$. At best we know that they are gravitational solutions with considerable entropy. Note that just below the energy $S_H$, at $\delta E = \frac{\pi}{\kappa_3^2}$, the specific heat of the subdominant stable configuration of supergravitons turns over and becomes negative, while a second, mechanically unstable configuration of self-gravitating radiation appears, as seen in \figref{fig:ads3_mass_star}, with positive specific heat but lower entropy than the stable star or the black hole.  
\item For $\frac{\pi}{\kappa^2} + \Delta = \frac{2\pi c}{3 L} + \Delta < \delta E$, a 3d black hole phase. For supersymmetric compactifications, such black holes are built on the Ramond ground state of the dual CFT.  There is a ``zero mass'' black hole, with vanishing horizon area, at energy $\delta E L = \frac{\pi L}{\kappa_3^2} = \frac{c}{12}$, where $G_N$ is the 3d Newton's constant. The entropy of the 3d black holes is:
\be
	S = \frac{12 \pi^2 L}{\kappa_3^2} \sqrt{ \frac{E \kappa_3^2}{\pi} - 1} = \frac{\pi c}{6} \sqrt{ \frac{12 \delta E L}{c} - 1}
\ee
This entropy starts at zero, so the 3d black hole phase must begin at an energy slightly higher than $\frac{\pi}{\kappa^2}$, at
$\frac{\Delta \kappa_3^2}{\pi} \sim {\cal O}\left(\frac{\kappa_3^{4/3}}{L^{2/3}}\right)$. Below this energy, the 3d black holes remain stable, merely entropically subdominant. Both the dynamically stable and dynamically unstable configurations of self-gravitating radiation persist up to $\delta E = \frac{1 + \pi}{\kappa^2}$, at which point the black hole is the only stable solution.
\end{enumerate}

For the D1-D5 system, we should be able to construct an order parameter $\phi$ again out of the square of a Polyakov loop.  The phases of the canonical ensemble are:
\begin{enumerate}
\item For $T < T_{HP} = \frac{1}{2\pi L}$, the dominant configuration is the supergravity gas, corresponding to region $A_L$ of the microcanonical phase curve.  Note that the spinodal point $S_H$ occurs at zero temperature: at this temperature, the ``zero mass'' black hole, with energy $\delta E = \frac{\pi}{\kappa_3^2}$ and vanishing entropy, first appears as a solution.  At finite energy below the Hawking-Page transition, we expect the self-gravitating gas and the BTZ black hole to be separated by a saddle point in the free energy; with the latter, metastable configuration corresponding to region $B_H$ on the microcanonical phase diagram. At arbitrarily low temperatures, there is no candidate ``small'' black hole.  However, the unstable equilibrium configurations of self-gravitating radiation, corresponding to the dashed line in \figref{fig:ads3_mass_star}, has the right order of magnitude of free energy (at zero temperature, it has the same free energy as the ``zero mass'' black hole).  The order parameter $\phi$ would appear to vanish, as it does for the stable minimum, since the bulk theory has no horizon. The combination of $\phi$ and the entropy of the configurations can distinguish the various solutions, however. Note that in this case, this saddle point in the free energy does not correspond to a micro canonically stable phase, nor is there any reason that it should.
\item At $T = T_{HP}$, the system undergoes a Hawking-Page transition, at which the BTZ black hole and the self-gravitating radiation gas has the same free energy. 
\item For $T > T_{HP}$, the BTZ black hole describes the global minimum of the free energy. The black hole corresponds to region $A_H$ of the microcanonical phase diagram, and the stable configuration of self-gravitating radiation  corresponds to region $B_L$.  At a temperature of this order, the black hole localized in six dimensions should appear as a saddle point with non vanishing $\phi$ separating the BTZ black hole from the metastable configuration of self-gravitating radiation; this should correspond to region $C$ of the microcanonical phase diagram.
\item At $T = T_{S_L} \sim \frac{1}{\ell_s} \sim \left(\frac{c}{6v}\right)^{1/4} \frac{1}{L}$, we reach the spinodal point $S_L$, at which point the 6d black hole goes through the correspondence point of \cite{Horowitz:1996nw}\ and a long string phase begins to dominate over a gas of self-gravitating radiation.  At this temperature, the unstable configuration of self-gravitating radiation should remain as a saddle point of the free energy. At higher temperatures, there is no metastable supergravity phase.
\end{enumerate}

\subsubsection{A different large-$c$ limit and the appearance of the Jeans instability}

As with self-gravitating radiation in $AdS_5$, we have found that a gas of supergravitons ceases to exist in $AdS_3$ for energies larger than $ML = \alpha c$ for some constant $\alpha$. 
These correspond to temperatures of order $T_J L \sim c^{1/6}\sim \left(\ell/\ell_p\right)^{1/6}$. We can again ask whether this will destabilize the metastable supergravity solution in the canonical ensemble at temperatures below the Hagedorn temperature. For the D1-D5 system, the Hagedorn temperature is:
\be
	T_H L \sim {L \over l_{s}} = \left(\frac{c}{6 v}\right)^{1/4}
\ee
Therefore $T_J \ll T_H$ when:
\be
	\frac{1}{g_6^2} = v \ll c^{1/3}
\ee
Small 6d string coupling $g_6$ corresponds to $v$ large, which is a strong coupling limit of the CFT. Nonetheless it is clearly possible to keep $g_6$ small and $c$ large consistent with this condition, so that bulk supergravity is a good approximation.  As before, in this limit the Hagedorn phase will disappear from the microcanonical phase diagram; the SUGRA gas will extend all the way to the point $S_L$.

\subsection{Implications for dynamics}

As we have seen, thermal AdS becomes unstable at sufficiently high temperature -- either at the string Hagedorn temperature \cite{Barbon:2001di,Kruczenski:2005pj}, or, with  an appropriate scaling limit, at a temperature parametrically below the string temperature and the Planck scale.  

This  fact has  interesting implications for the question of unitarity of black holes.  In AdS, large black holes are thermodynamically stable---they do not evaporate.  However, as was pointed out in \cite{Maldacena:2001kr}, semi-classically they still appear to violate unitarity in a very precise sense.  Bulk correlation functions computed in black hole backgrounds decay exponentially at late times.  For example, the typical behavior of a response function at late times is
\be \label{fdef}
f(t) \equiv |\langle O(0) O(t) \rangle|^2 \sim e^{-a t/\beta },
\ee
where $a$ is a numerical constant and $O$ is some (suitably regulated) bulk operator  \cite{Dyson:2002pf,Barbon:2003aq}.  This exponential decay can be understood as arising from the tendency of horizons to swallow whatever is thrown into them, with a response that is quasi normal (i.e. the frequency has an imaginary part).  Hence all trace of any initial perturbation is lost after a few thermal timescales.  

By contrast, a finite temperature quantum system with finite entropy and unitary time evolution cannot behave this way.  The long time average of correlators is non-zero:
\be \label{timeav}
{1 \over 2 T} \int_{-T}^{T} f(t) dt \sim e^{-b S}
\ee
for some constant $b$ -- inconsistent with \eqref{fdef}.  In addition there are Poincare recurrences, meaning that $f(t)$ is a quasiperiodic function of time that returns arbitrarily close to its initial value infinitely many times. \cite{Dyson:2002pf,Barbon:2003aq} This inconsistency is a particularly precise version of the black hole information paradox.

In \cite{Maldacena:2001kr}, Maldacena proposed that the non-zero time-average could be accounted for by incorporating other geometries besides the black hole in the calculation of the bulk correlation functions.  Specifically, he pointed out that thermal AdS at the Hawking temperature has the same asymptotics as the black hole, but (due to its lack of a horizon) correlators computed in it oscillate rather than decay.  Furthermore, in the high-temperature phase where the black hole is thermodynamically dominant, the contribution of thermal AdS should  be suppressed by a relative factor of 
\be \label{Fdiff}
e^{-\beta(F_{AdS}-F_{BH})} \sim e^{-c S},
\ee
where $c$ is an ${\cal O}(1)$ constant.  Since correlators in the black hole decay exponentially, the long-time average would come only from the oscillating thermal AdS contribution, and hence have the correct exponential-in-$S$ suppression.  The issue of origin of the Poincare recurrences would still remain \cite{Barbon:2003aq}, but perhaps that could arise from quantum interference in the integral over geometries.

However, the instability of thermal AdS at either the Hagedorn temperature $T_H$ (in the usual large-N/large-c limit) or the Jeans temperature $T_J$ means that above these temperatures, the proposal in \cite{Maldacena:2001kr}\ fails even at the level of the long-time average  \cite{Barbon:2003aq}. In particular, in the limits described in \S3.2.2\ and \S3.3.2, thermal AdS becomes perturbatively unstable at a temperature where gravity is still a good description, so that the free energy in \eqref{Fdiff} becomes imaginary and thermal AdS can no longer account for the non-zero time-average in \eqref{timeav}.  Of course it is always possible that there is yet another geometry or stringy solution that is not captured by our analysis and could replace AdS at  high temperatures (for instance, a star solution that is inhomogeneous on the sphere, or a solution with a sufficiently non-linear equation of state), but we are not aware of any candidate.



\section*{Acknowledgements}
It is a pleasure to thank Bulbul Chakraborty, Stanley Deser, Matthew Headrick, Gary Horowitz, Matthew Johnson, William Klein, Don Marolf, John McGreevy, Dhritiman Nandan, Massimo Porrati, Nathaniel Reden, Stephen Shenker, Hugo Touchette, Robert Wald, and Andrew Waldron for useful conversations (some from many years ago).  The work of MK is supported in part by the NSF through grant PHY-1214302, and that of MK and SS was supported by the John Templeton Foundation. The work of Albion Lawrence is supported in part by the DOE via award DE-SC0009987. MMR is supported by the Kadanoff Center for Theoretical Physics. The opinions expressed in this publication are those of the authors and do not necessarily reflect the views of the John Templeton Foundation.

\appendix

\section{Black hole solutions and thermodynamic quantities}
\label{app:conventions}

\subsection{$AdS_5$ black holes}

In this Appendix we recall the black hole solutions in $AdS_5$ and their energy and temperature, expressed in terms of both bulk parameters and, for the case that the dual gauge theory is ${\cal N}=4$ large-N super-Yang-Mills, in terms of the gauge theory parameters.

The Hawking-Page black hole in five dimensions is \cite{Hawking:1982dh,Witten:1998qj,Witten:1998zw,Horowitz:1998ha}:
\begin{eqnarray}
	ds^2 & = & - f(r) dt^2 + \frac{dr^2}{f(r)} + r^2 d\Omega_3^2\nonumber\\
	f(r) & = & 1 + \frac{r^2}{L^2} - \frac{r_0^2}{r^2}
\end{eqnarray}
where $d\Omega_3^2$ is the volume element on the unit three-sphere, and $L$ is the radius of curvature of the ambient $AdS_5$ spacetime.  The energy of this configuration, measured with respect to vacuum $AdS_5$, is
\be
	E = \frac{3 \pi^2 r_0^2}{\kappa_5^2} = \frac{3}{4} N^2 \frac{r_0^2}{L^3}
\ee
where $\kappa_5^2 = 8\pi G_{N,5}$ is the 5d gravitational coupling constant, and the rank $N$ of the dual gauge theory is determined by:
\be
	N^2 = \frac{4\pi^2 L^3}{\kappa_5^2} 
\ee
The black hole horizon occurs at the largest zero of $f(r)$, denoted $r_+$:
\be
	r_+^2 = \frac{-L^2 + \sqrt{L^4 + 4L^2r_0^2}}{2}
\ee
The temperature is
\be
	T = \frac{r_+}{\pi L^2} + \frac{1}{2\pi r_+}\ ,
\ee
and the entropy is
\be
	S = \frac{4\pi^3 r_+^3}{\kappa_5^2} = \pi N^2 \left(\frac{r_+}{L}\right)^3
\ee

We note two particularly interesting points in this family of solutions. The Hawking-Page transition, above which the black hole dominates the thermodynamics of string theory on $AdS_5\times S_5$, happens at $r_+ = L$, $T = 3/(2\pi L)$.  The energy of the black hole at this point is $E_{HP} L = \frac{6 \pi^2 L^3}{\kappa_5^2} = \frac{3}{2} N^2$. Next, the temperature below which the black hole has negative specific heat, corresponding to the spinodal point $S_H$ in \figref{fig:FirstOrderNE}, occurs when $r_+ = \frac{L}{\sqrt{2}}$, $T_{S_H} = \frac{\sqrt{2}}{\pi L}$, $E_{S_H}L = \frac{9}{16} N^2$. Finally, an instability for the black hole to localize along the $S_5$ is believed to set in at $r_+ \sim .4259 L$, at a mass and temperature below that of $E_{S_H}$, $T_{S_H}$.

\subsection{Black holes in ${\mathbb R}^{1,9}$}

The solutions for black holes localized in ten dimensions in $AdS_5\times S_5$ are not known.  Here we will consider 10d Schwarzschild black holes in flat space in the limit that their Schwarzschild radius is much smaller than the AdS radius $L$. 
The full solution is
\begin{eqnarray}
	ds^2 & = & - f(r) dt^2 + \frac{dr^2}{f(r)} + r^2 d\Omega_8^2 \nonumber\\
	f(r) & = & 1 - \left(\frac{r_0}{r}\right)^7
\end{eqnarray}
The energy is
\be
	E = \frac{256 \pi^4 r_0^7}{105 \kappa_10^2} = \frac{64}{105} N^2 \frac{r_0^7}{L^8}
\ee
where $\kappa_10^2$ is the ten-dimensional gravitational coupling.  In the $AdS_5\times S_5$ background dual to ${\cal N}=4$ super-Yang-Mills, the rank of the dual gauge group is:
\be
	N^2 = \frac{4 \pi^4 L^8}{\kappa_{10}^2}
\ee
The entropy of the 10d black hole is
\be
	S = \frac{64 \pi^5 r_0^8}{105 \kappa_10^2} = \frac{16 \pi}{105} N^2 \left(\frac{r_0}{L}\right)^8
\ee
Finally, the temperature is
\be
	T = \frac{7}{4\pi r_0}
\ee

\subsection{$AdS_3$ black holes}

The metric for the 3d BTZ black hole is \cite{Banados:1992wn,Banados:1992gq}:
\begin{eqnarray}
	ds^2 & = & - f(r) dt^2 + \frac{dr^2}{f(r)} + r^2 d\phi^2\nonumber\\
	f(r) & = & \frac{r^2 - r_+^2}{L^2}
\end{eqnarray}
Here $\phi \equiv \phi + 2\pi$, and $L$ is the radius of curvature of the ambient $AdS_3$ space-time.  Note that at $r_+ = 0$, this black hole is a compactification of $AdS_3$ in Poincar\'e coordinates. It has a null singularity along the Poincar\'e horizon $r = 0$, and corresponds to the zero-temperature black hole. In $2+1$ dimensions, the BTZ black hole always has positive specific heat.

The energy of the BTZ black hole is
\be
	E = \frac{\pi}{\kappa_3^2} \left(1 + \frac{r_+^2}{L^2}\right) = \frac{c}{12 L} \left(1 + \frac{r_+^2}{L^2}\right)
\ee
in units in which the $AdS_3$ vacuum has energy $- \pi/\kappa_3^2 = - c/(12 L)$. When the bulk gravitational theory has a $2d$ CFT dual, $c = 12 \pi L/\kappa_3^2$ is the central charge. The entropy is
\be
	S = \frac{2\pi^2 r_+^2}{\kappa_3^2} = \frac{\pi c r_+^2}{6 L^2}
\ee
and the temperature is
\be
	T = \frac{r_+}{\pi L^2}
\ee
The Hawking-Page transition occurs at
\be
	r_+ = \frac{L}{2}
\ee
or at $T = \frac{1}{2\pi L}$, $E = \frac{5 c}{48 L}$, $S = \frac{\pi c}{12}$.

\subsection{Black holes in ${\mathbb R}^{1,5}$}

We will be interested in black holed localized in six dimensions, in the case of $AdS_3\times S_3$ compactifications.  As with the discussion of $10d$ black holes above, we will consider Schwarschild black holes in flat space, approximating black holes whose horizon radius is smaller than the radius of curvature of $AdS_3\times S_3$.

The metric is
\begin{eqnarray}
	ds^2 & = & - f(r) dt^2 + \frac{dr^2}{f(r)} + r^2 d\Omega_4^2\nonumber\\
	f(r) & = & 1 - \left(\frac{r_0}{r}\right)^3
\end{eqnarray}
where $d\Omega_4$ is the volume element of the unit $4$-sphere. The mass of this black hole is
\be
	E = \frac{16 \pi^2 r_0^3}{3 \kappa_6^2} = \frac{2}{9\pi} c \frac{r_0^3}{L^4}
\ee
where, when the theory has a $2d$ CFT dual, the central charge is
\be
	c = \frac{24 \pi^3 L^4}{\kappa_6^2}
\ee
The entropy is
\be
	S = \frac{16 \pi^3 r_0^4}{3 \kappa_6^2} = \frac{2}{9} c \frac{r_0^4}{L^4}
\ee
and the temperature is
\be
	T = \frac{3}{4 \pi r_0}
\ee

\section{The perturbation equations for $AdS_{d+1}$}
\label{app:full_pert}

In this appendix we give the full system of equations for s-wave perturbations of a fluid in $AdS_{d+1}$, without any assumption on the equation of state. We will continue to work in the gauge where $r$ measures the size of the $S_{d-1}$, so $ds^2_{r,t~\mathrm{ const.}}=r^2d\Omega_{d-1}$.  As we are considering only s-wave perturbations, we include the perturbations $\delta g_{tt},~\delta g_{rr},~\delta u^r,~\delta u^t,\delta\rho \propto e^{-i\omega t}$. Identifying $u^a$ as the fluid velocity means we need $u^2 = -1+\Op(\delta^2)$, which gives an algebraic constraint 
\be
\delta u^t = \frac{e^{3\chi/2} }{2 f^{3/2}}\delta g_{tt}.
\ee
We are then left with the perturbative conservation and Einstein equations. We find appropriate linear combinations to solve for $\delta g_{rr},~\delta g_{rr}',\delta g_{tt}',\delta g_{tt}'',\delta {u^r}',~\delta\rho'$. Our symmetries and gauge constraints have given us an algebraic constraint for $g_{rr}$ and $\delta u^r$,
\be
\delta g_{rr} + \frac{2i \kappa^2  r e^{\chi/2}(p+\rho)}{\omega f^{3/2} (d-1)}\delta u^r=0.
\ee
It is beneficial to use this to remove $\delta g_{rr}$ from the remaining equations. We find the following equations for $\delta g_{tt}',~\delta {u^r}',~\delta\rho'$: 
\be
\delta g_{tt}' =
-\frac{d-2}{r} \delta g_{tt}-\frac{2 \kappa^2 r e^{-\chi}p_{,\rho}}{d-1}\delta\rho\nonumber\ee
\be
+\left(\frac{d r}{L^2f}+\frac{d-2}{r f}+\frac{2\kappa^2 r p}{(d-1)f}\right)\left(\delta g_{tt}+\frac{2i\kappa^2 r e^{-3\chi/2}\sqrt{f}(p+\rho)}{(d-1)\omega}\delta u^r \right)
\ee
\be
\delta\rho'=-\left( \frac{\frac{d r}{2L^2f}+\frac{\kappa^2 r p}{(d-1)f}}{p_{,\rho}}+\frac{r\kappa^2(2p+\rho)}{(d-1)f}+\frac{(d-2)(1-f)(1+p_{,\rho})}{2rf p_{,\rho}}\right.\nonumber\ee
\be
\left.
+\frac{d r}{2L^2f}-\frac{p+\rho}{2L^2rfp_{,\rho}^2}\left\{d r^2 + (d-2)L^2(1-f)+\frac{2\kappa^2L^2r^2p}{d-1} \right\}p_{,\rho\rho}
\right)\delta\rho\nonumber
\ee
\be
+\left(
\frac{i\kappa^2r^2e^{-\chi/2}(p+\rho)^2}{(d-1)f^{3/2}p_{,\rho}}\left\{ 
\frac{d}{L^2}+\frac{d-2}{r^2}+\frac{2\kappa^2 p}{d-1}
\right\}
+\frac{i\omega e^{\chi/2}(p+\rho)}{p_{,\rho} f^{3/2}}
 \right)\delta u^r,
\ee
\be
\delta {v^r}'=\frac{i\omega e^{\chi/2}}{\sqrt{f} (p+\rho)}\delta\rho
-
\left(
\frac{d-1}{r}+\frac{d-2}{2rp_{,\rho}}-\frac{r}{2fp_{,\rho}}\left\{ 
\frac{d}{L^2}+\frac{d-2}{r^2}+\frac{2\kappa^2 p}{d-1}
\right\}
\right)\delta u^r.
\ee

Note that we have used the constraints to remove the metric perturbations from the equations for $\delta\rho,~\delta u^r$.

\eject
\bibliographystyle{klebphys2}
\bibliography{instabrefs2,mmr_refs}

\vskip .5cm
\vskip .5 cm

\end{document}